\documentclass[12pt, draftclsnofoot, onecolumn]{IEEEtran}

\usepackage{cite}
\usepackage{graphicx}
\usepackage{times}
\usepackage{latexsym}

\usepackage[mathscr]{eucal}
\usepackage[center]{caption2}
\usepackage{array}
\usepackage{fancyhdr}
\usepackage{amsmath,amssymb}
\usepackage{bm} 
\usepackage{subfigure}
\usepackage{multirow}

\ifCLASSINFOpdf
\else
\fi

\newtheorem{de}{Definition}
\newtheorem{thm}{Theorem}

\newtheorem{rem}{Remark}

\newcommand{\defeq}{\stackrel{\Delta}{=}}

\newcommand{\qed}{\hfill\ensuremath{\blacksquare}}

\newcommand{\nn}{\nonumber}

\newcommand{\vect}[1]{{\lowercase{\mathbf{#1}}}}
\newcommand{\mat}[1]{{\uppercase{\mathbf{#1}}}}

\newcommand{\tr}{{\rm{tr}}}

\renewcommand{\d}{\vect{d}} 
\newcommand{\e}{\vect{e}}

\newcommand{\h}{\vect{h}}

\newcommand{\n}{\vect{n}}

\renewcommand{\r}{\vect{r}}

\renewcommand{\v}{\vect{v}} 
\newcommand{\w}{\vect{w}}
\newcommand{\x}{\vect{x}}

\newcommand{\A}{\mat{A}}
\newcommand{\B}{\mat{B}}
\newcommand{\C}{\mat{C}}

\renewcommand{\H}{\mat{H}} 
\newcommand{\I}{\mat{I}}

\newcommand{\Q}{\mat{Q}}
\newcommand{\R}{\mat{R}}

\newcommand{\Lc}{{\cal L}}

\newcommand{\Qc}{{\cal Q}}
\newcommand{\Rc}{{\cal R}}

\newcommand{\Rt}{{\tilde \R}}

\newcommand{\Ra}{{\bar R}}

\newcommand{\Cb}{{\mathbb C}}

\newcommand{\Eb}{{\mathbb E}}

\newcommand{\Rb}{{\mathbb R}}

\newcommand{\Lambdam}{\hbox{\boldmath$\Lambda$}}

\usepackage{xcolor}

\makeatother

\usepackage{algpseudocode}
\usepackage{algorithm}
\usepackage{amsfonts}

\newcommand{\gt}{\tilde {g}}

\newcommand{\tsum}{\text{sum}}

\newcommand{\tub}{\text{ub}}
\newcommand{\tlb}{\text{lb}}
\newcommand{\rank}{\text{rank}}

\newcommand{\tst}{\mbox{s.t. }}
\newcommand{\bb}{{\bf{0}}}

\newcommand{\hht}{\tilde {\bf h}}

\graphicspath{{Figure/}}

\begin{document}
\title{Beam Domain Massive MIMO for Optical Wireless Communications with Transmit Lens}

\author{
Chen~Sun, Xiqi~Gao, Jiaheng~Wang, Zhi~Ding, and Xiang-Gen~Xia
\thanks{C. Sun, X. Q. Gao, and J. Wang are with the National Mobile Communications Research Laboratory,
Southeast University, Nanjing, 210096, China. X. Q. Gao is the correspondence author (email: \{sunchen,xqgao,jhwang\}@seu.edu.cn).}
\thanks{Z. Ding is with the Department of Electrical and Computer Engineering, University of California, Davis,
CA 95616, USA (email: zding@ucdavis.edu). }
\thanks{X.-G. Xia is with the Department of Electrical and Computer Engineering,
University of Delaware, Newark, DE 19716, USA (e-mail: xxia@ee.udel.edu).}
}
 
\maketitle

\begin{abstract}
This paper presents a novel massive multiple-input multiple-output (MIMO) transmission in beam domain for optical wireless communications. The optical base station equipped with massive 
optical transmitters communicates with a number of user terminals (UTs) through a transmit lens. 
Focusing on LED transmitters, we analyze light refraction of the lens and establish a channel model for optical massive MIMO transmissions.
For a large number of LEDs,
channel vectors of different UTs become asymptotically orthogonal.
We investigate the maximum ratio transmission and regularized zero-forcing precoding
in the optical massive MIMO system,
and propose a linear precoding design to maximize the sum rate. 
We further design the precoding when the number of transmitters grows asymptotically large, and show that beam division multiple access (BDMA) transmission achieves the asymptotically optimal performance for sum rate maximization. 
Unlike optical MIMO without a transmit lens, BDMA can increase the sum rate proportionally to $2K$ and $K$ under the total and per transmitter power constraints, respectively, where $K$ is the number of UTs.
In the non-asymptotic case, we prove the orthogonality conditions of the optimal power allocation in beam domain and propose efficient beam allocation algorithms. Numerical results confirm the significantly improved performance of our proposed beam domain optical massive MIMO communication approaches.

\end{abstract}

\begin{IEEEkeywords}
Optical massive MIMO communication,
transmit lens,
Beam division multiple access (BDMA).
\end{IEEEkeywords}

\newpage

\section{Introduction} 
\label{sec:introduction}

Optical wireless communication systems rely on optical radiations to transmit information
with wavelengths ranging from infrared to ultraviolet \cite{Elgala2011Indoor}.
Base station (BS) commonly employs optical transmitters, such as
light-emitting diodes (LEDs), to convert the electrical signals to optical signals.
Recently, laser diodes (LDs) are considered as potential
sources for optical communication due to high modulation bandwidth, efficiency, and
beam convergence \cite{7842427}.
User terminals (UTs) employ photodetectors like photodiodes as optical receivers to convert the optical power into electrical current.
Optical communications can significantly relieve the crowed radio frequency (RF) spectrum,
provide high speed data transmission \cite{Azhar2010Demonstration},
and achieve simple and low-cost modulation and demodulation through intensity modulation
and direct detection (IM/DD) \cite{Gao2015DD}.
Thus, optical wireless communication has attracted increasing attention from both academia and industry \cite{Fath2013Performance,Kedar2004Urban,Burton2014Experimental}.

To achieve high data rate in optical communications,
multiple separate LED/LD arrays are usually utilized in the BS to provide higher data rate by means of spatial multiplexing. 
As a result, multiple-input multiple-output (MIMO) technique is a natural progression for optical communication systems \cite{Zeng2009High,Burton2014Experimental}.
As the nature of optical downlink communication is broadcast network, multiple UTs should be well supported.
BSs simultaneously transmit signals to all UTs, resulting in the so-called multi-user interference, consequently degrades the performance.
Thus, multi-user MIMO (MU-MIMO) has been studied and several precoding schemes have been proposed, which are different from conventional
RF systems since only real-valued non-negative signals can be transmitted 
\cite{6638582,6825136,Hong2013Performance,7124415,8113500,Wang:16,7254237}.
In \cite{6638582}, the performances of zero forcing and dirty paper coding schemes were compared. 
An optimal linear precoding transmitter was derived based on the minimum
mean-squared error (MMSE) criterion in \cite{6825136}, while block diagonalization precoding algorithm was investigated
in \cite{Hong2013Performance}. 
The work in \cite{7124415} considered the multiuser transceiver design under per-LED optical power constraints.
Moreover, the biased multi-LED beamforming \cite{8113500}
and transmit designs in orthogonal frequency-division multiplexing (OFDM) systems \cite{Wang:16,7254237} were investigated.
Furthermore, recently proposed massive MIMO with tens or hundreds of antennas in RF systems \cite{Marzetta}
was applied into optical communication systems \cite{Gao2015Low} to increase spectrum efficiency.

Since optical communication systems employ intensity modulation and direct detection
and line-of-sight (LOS) scenario is mostly considered,
highly correlated channels limit the system performance \cite{Zeng2009High}.
Imaging receiver is employed to separate signals from different directions, which was originally proposed for infrared communications \cite{200761}.
In \cite{Zeng2009High}, the authors investigated non-imaging and imaging MIMO optical communications, 
and indicated that the imaging receiver can potentially offer higher spatial diversity. 
Due to poor imaging quality leading to interference from different LED arrays,
some works proposed an imaging receiver with a fisheye lens \cite{6891135}
or a hemispherical lens \cite{6497451} to provide high-spatial diversity for MIMO signals.
Many works study the receive lens to separate the signals from different LED arrays.
However, there are no work considering a transmit lens at the BS.
Without a transmit lens, each transmitter, such as an LED, is omni-directional.
Since the distance between the LED array and a UT is much larger than the LED array size,
the channels between different transmitters and UT are highly correlated \cite{Butala2014Performance}.
Thus, one LED array transmits only one data stream \cite{Zeng2009High}.
To support multiple users, multiple separate LED arrays are required.
The number of LED arrays dominates the number of served UTs.

Motivated by the receive lens to separate lights from different LED arrays, we employ a transmit lens to refract the lights from different transmitters in an array towards different directions, providing high spatial resolutions.
In this paper, we study beam domain optical wireless massive MIMO communications, where a
BS equipped with a large number of optical transmitters serves a number of UTs simultaneously through a transmit lens.
The fundamental principle of the transmit lens is to provide variable refraction angles so as to achieve the angle-dependent energy focusing property.
Specifically, lights from different transmitters are sufficiently separated by the transmit lens.
We focus on LED transmitters as an example, while the proposed scheme can be applied to LD transmitters or optical fiber ports connected to optical transceivers.
{
We first establish the optical channel model with a transmit lens,
and analyze the relationship between the emitted light and refractive light.
The refractive lights from one LED are concentrated within a small angle and thus generate a narrow beam.
More interestingly, for a large number of LEDs, the channel vectors of different UTs become asymptotically orthogonal,
implying that the optical massive MIMO system has the potential to simultaneously serve a number of UTs.}

Based on the established optical channel model, we investigate linear transmit design
including maximum ratio transmission (MRT) and regularized zero-forcing (RZF),
and propose a linear precoding design for the sum rate maximization under the total and per LED power constraints. 
Moreover, we investigate the precoding design when the number of LEDs goes to infinity, and 
obtain important insights.
The asymptotically optimal transmission is that
the transmit beams for different UTs are non-overlapping,
and that different transmitters send independent signals to different UTs.
Therefore, beam division multiple access (BDMA) transmission can achieve the asymptotically optimal performance.
Compared with the conventional transmission without a transmit lens,
the BDMA transmission increases the sum rate proportionally to $2K$ ($K$ is the number of UTs) and $K$ under the total power and per LED power constraints, respectively.
For a limited number of LEDs, we consider the beam domain power allocation,
prove the orthogonality of the optimal power allocation, and provide efficient beam allocation algorithms.
Numerical results illustrate significant performance gains of our proposed optical massive MIMO communication approaches.

We adopt the following notations throughout this paper:
Upper (lower) bold-face letters denote matrices (column vectors). ${\bf {I}}$ denotes
the identity matrix, while ${\bf {1}}$ denotes an all-one matrix and $\bb$ denotes zero matrix.
Numeral subscripts of matrices and vectors, if needed, represent their sizes.
Also, matrix superscript $(\cdot)^T$ denotes matrix transpose.
We use  $\tr(\cdot)$ and $\det(\cdot)$ to represent matrix trace and determinant operations, respectively.
The inequality $\A\succeq {\bf 0}$ denotes a positive semi-definite Hermitian matrix $\A$.
We use $[{\bf A}]_{mn}$ to denote the $(m,n)$-th element of matrix $\bf A$.

\section{Transmit Lens Based Optical Massive MIMO System and Channel Modeling} 
\label{sec:system_model}

\subsection{Optical Massive MIMO System with Transmit Lens} 
\label{sub:system_setup}

\begin{figure}[htbp]
	\centering
	{\includegraphics[width = 0.75\textwidth]{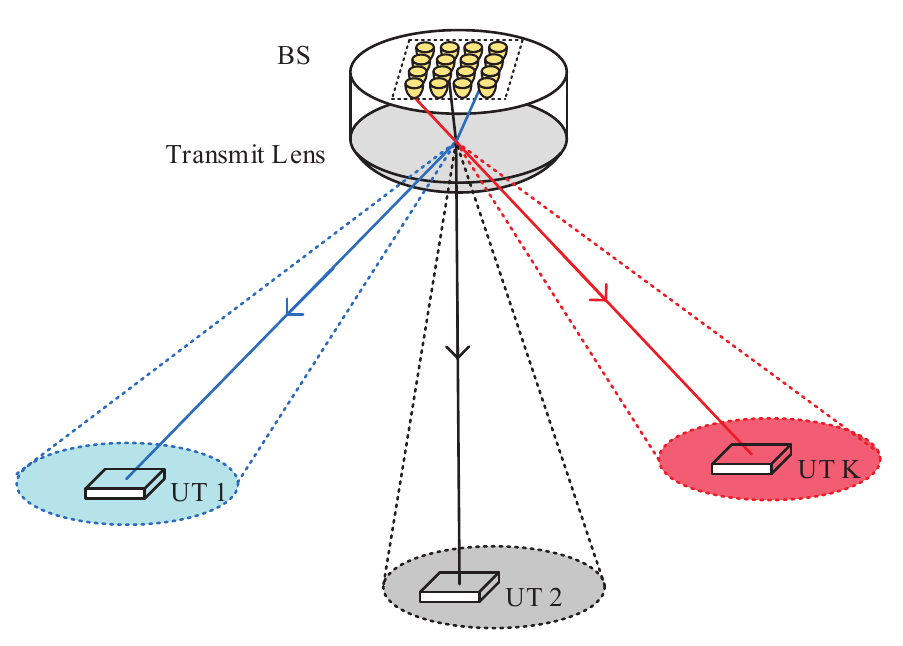}}
	\caption{Optical MIMO system with a transmit lens. }\label{fig:F1}
\end{figure}

We consider an optical massive MIMO system consisting of a BS, equipped with $M^2$ transmitters and a transmit lens,
and $K$ UTs, each of which has one photodetector employed as a receiver.
The transmitters can be LEDs, LDs, or optical fiber ports connected to optical transceivers. In this paper, we focus on LED transmitters, {and each LED emits lights with the same wavelength.}
{We can similarly design the optical system according to the radiation pattern of the LD transmitter \cite{7842427}.}
With a transmit lens at the BS, the lights emitted from different LEDs passing through the lens are refracted to different directions, as shown in Fig.~\ref{fig:F1}.

BS employs the square LED array to transmit signals to UTs.
Denote $\x_k\in \Rb^{M^2 \times 1 }$ as the signal intended for the $k$th UT,
and the received signal at the $k$th UT can be written as
\begin{align}\label{eq:broadcast}
    y_k = \h_k^T \x + z_k  
          = \h_k^T \x_k + \h_k^T \left( \sum_{k' \neq k } \x_{k'} \right) + z_k  ,      
\end{align}
where $\x = \sum_k \x_k$ is the summation of all signals,
$\h_k^T\in\Rb^{1\times M^2}$ is the channel vector from the LEDs to the $k$th UT,
and $z_k$ is the noise at the receiver, which contains ambient-induced shot noise and thermal noise,
and is generally modeled as a real-valued additive white Gaussian variable with zero mean and variance $\sigma^2$ \cite{1277847}.
Here, without loss of generality, we assume a unit noise variance (i.e., $\sigma^2=1$).

\subsection{Refraction of Light Passing Through Lens} 
\label{sub:physical_channel_model_with_transmit_lens}


\begin{figure}[!h]
\centering
\subfigure[]{\includegraphics[width=0.35\textwidth]{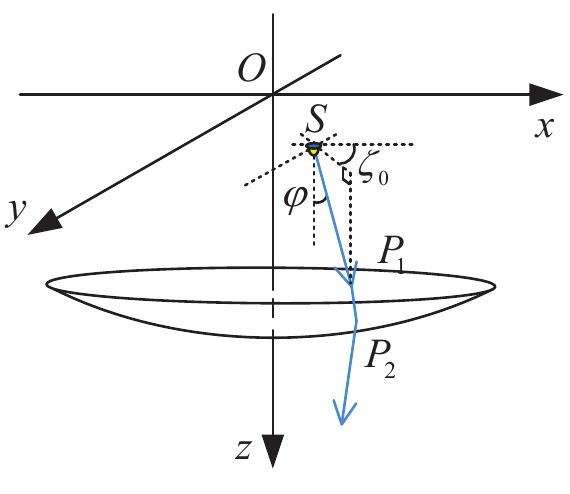}\label{fig:Lens3a:0}}
\subfigure[]{\includegraphics[width=0.35\textwidth]{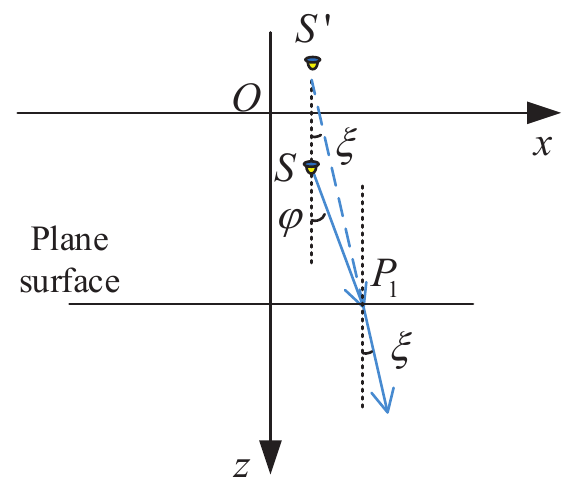}\label{fig:Lens3a:1}}
\subfigure[]{\includegraphics[width=0.35\textwidth]{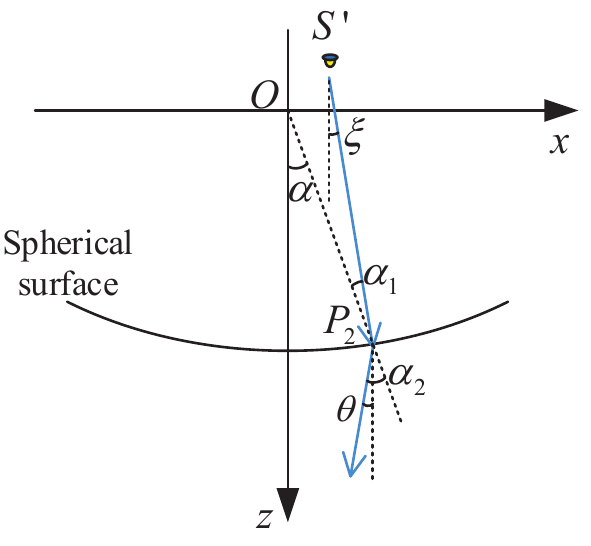}\label{fig:Lens3a:2}}
\subfigure[]{\includegraphics[width=0.35\textwidth]{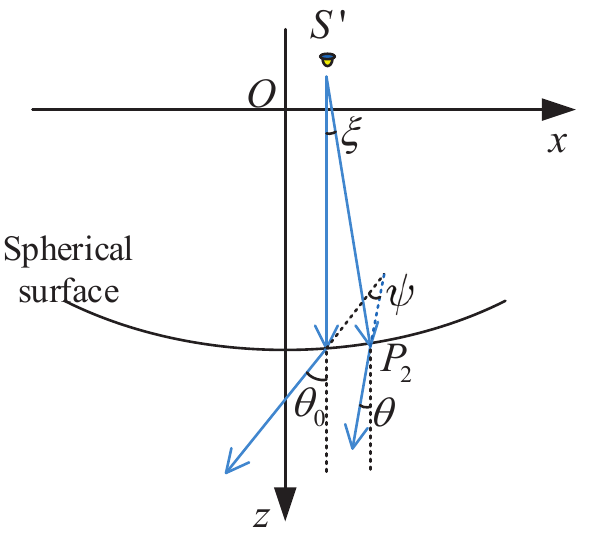}\label{fig:Lens3a:3}}
\caption{(a) Illustration of a light passing through the lens; (b) refraction at the plane surface; (c) refraction at the spherical surface; and (d) refractive angle with respect to the center light.}\label{fig:Lens3a}
\end{figure}

{
We first analyze the refraction of the light from one LED passing through the lens.
Suppose that the $(i,j)$th LED in the square LED array is located at point $S$, with coordinates $(x_S,y_S,z_S)$, as shown in Fig.~\ref{fig:Lens3a:0}.
For simplicity, we omit the subscript $(i,j)$.
Let $\varphi_C$ denote the semi-angle of half intensity of the LED.
The luminous intensity of an LED generally follows the Lambertian radiant distribution \cite{Zeng2009High}.
Specifically, for the light with the polar angle $\varphi$, the luminous intensity can be expressed as
\begin{align}\label{eq:I_0}
    I_0(\varphi) = \frac{m_L+1}{2\pi} \cos^{m_L} (\varphi),
\end{align}
where the order of Lambertian emission is given by $m_L = -\log 2 / \log(\cos(\varphi_{C}))$.


We exploit a spherical lens, with the refractive index $n$, in front of the LED.
The plane surface is facing the LED, and the center of the spherical surface,
whose radius is $R$, is the origin of the orthogonal coordinate system,
as shown in Fig.~\ref{fig:Lens3a}.
Consider a light emitted from the LED with the polar angle $\varphi$ and the azimuthal angle $\zeta_0$.
The light is first refracted through the planar surface at $P_1$, as shown in Fig.~\ref{fig:Lens3a:1},
and refracted through the spherical surface at $P_2$, as shown in Fig.~\ref{fig:Lens3a:2}.

At $P_1$, the incident angle is $\varphi$,
and the refractive angle is $\xi$.
It follows from Snell's law of $\sin \varphi / \sin \xi = n$.
The refractive light can be treated as light emitted from a virtual light source $S'$ $(x_{S},y_{S},z_{S'})$ with the polar angle $\xi$ and the azimuthal angle $\zeta_0$.

Next, we analyze the light passing through the spherical surface at $P_2$.
Denote the unit vector of the incident light at $P_2$ by
$\v = (\sin(\xi)\cos(\zeta_0),\sin(\xi)\sin(\zeta_0),\cos(\xi))$.
The unit normal vector $\n$ at $P_2=(x_2,y_2,z_2)$ is
$\n = \left( -{x_2}/{R},-{y_2}/{R},-{z_2}/{R}  \right)$.
According to Snell's law, 
the unit vector $\r$ of the refractive light at $P_2$ is given by
\begin{align}\label{eq:r}
    \r = n \v + (n c - \sqrt{1-n^2(1-c^2)})\n ,
\end{align}
where 
\begin{align}
    c = -\n \cdot \v=\frac{x_2}{R}\sin(\xi)\cos(\zeta_0)+\frac{y_2}{R}\sin(\xi)\sin(\zeta_0)+\frac{z_2}{R}\cos(\xi).
\end{align}
Denote the polar angle of the refractive light as $\theta$
and the azimuthal angle as $\zeta_1$.
From \eqref{eq:r}, we have
\begin{align}
    \cos(\theta) &= n\cos(\xi) - (n c - \sqrt{1-n^2(1-c^2)})\frac{z_2}{R},  \label{eq:rz} \\
    \sin(\theta) \cos(\zeta_1)&= n\sin(\xi)\cos(\zeta_0) - (n c - \sqrt{1-n^2(1-c^2)})\frac{x_2}{R} , \label{eq:rx} \\
    \sin(\theta) \sin(\zeta_1) &= n\sin(\xi)\sin(\zeta_0) - (n c - \sqrt{1-n^2(1-c^2)})\frac{y_2}{R}.  \label{eq:ry}    
\end{align}
The exact relationship between
the incident light with angle $(\varphi,\zeta_0)$ and the refractive light with angle $(\theta,\zeta_1)$ is characterized by equations \eqref{eq:rz}--\eqref{eq:ry},
which, however, are quite involved and offer no clear insight.
Thus, we would like to establish a simpler relationship via proper approximations
in geometrical optics \cite{hecht2002optics}.

As the luminous intensity of the LED exhibits rotational symmetry,
we focus on the relationship between $\theta$ and $\varphi$,
and consider the $S'OP_2$ plane in Fig.~\ref{fig:Lens3a:2} and \ref{fig:Lens3a:3}.
Denote the angles of incidence and refraction at $P_2$ by $\alpha_1$ and $\alpha_2$, respectively,
and the angle of the normal by $\alpha$.
Then, $\theta$ can be expressed as 
\begin{align}
    \theta = \alpha_2 - \alpha = \alpha_2 - \xi - \alpha_1.
\end{align}
From Snell's law, $\sin(\alpha_2) = n \sin(\alpha_1)$ and $\sin(\varphi) = n \sin(\xi)$.
Given small values of $\varphi$, $\xi$, $\alpha_1$ and $\alpha_2$,
we have $\sin(\alpha)\approx \alpha$\footnote{{
    This approximation does not need the angle to be very small.
    When the incident angle is $30^\circ$ (0.52 rad), the value of sine function is 0.5, and the approximation remains good.
}}.
Thus,
$\theta$ can be expressed as
\begin{align}
    \theta \approx (n-1)\alpha - \varphi.
\end{align}
Meanwhile, we have $\alpha \approx (\sqrt{x_S^2+y_S^2} + \varphi(R-z_{S'})/n)/R$ and $(z_p-z_{S'}) \approx n(z_p-z_S)$,
where $z_p$ is the position of the plane surface. Consequently, $\theta$ can be approximated by
\begin{align}\label{eq:varphi2}
    \theta &\approx  \underbrace{ (n-1) \frac{\sqrt{x_S^2+y_S^2}}{R}}_{\theta_0} - 
    \underbrace{ \frac{\varphi}{n} \left( 1 + \frac{n z_S}{\frac{R}{n-1}} - \frac{(n-1)^2 z_p}{R} \right)}_{\psi}.
\end{align}
\begin{rem}
As revealed by \eqref{eq:varphi2}, the polar angle of the refractive light $\theta$, as shown in Fig. 2(c),
can be divided into two parts. 
The first term $\theta_0$ represents the refractive angle of the center light (i.e., $\varphi=0^\circ$) and
depends on the horizontal position of the LED on the $XOY$ plane. 
The second term $\psi$ represents the angle between the refractive light and the refractive center light,
and is only determined by the vertical positions of the LED and the lens.
Thus, the refractive light shows rotational symmetry with respect to the center light.
For hemispherical lenses, we have $z_p = 0$ and
\begin{align}\label{eq:hemispherical}
    \theta &\approx \theta_0 - {\varphi} \left( {1}/{n} + (n-1){z_S}/R \right).
\end{align}
For thin lenses, $z_p$ approaches $R$, leading to
\begin{align}\label{eq:thin}
    \theta &\approx \theta_0 - {\varphi} \left( 2-n + (n-1){z_{S}}/R  \right),
\end{align}
where ${R}/{(n-1)}$ is the focal length $f$ \cite{hecht2002optics}.
\end{rem}

Now, we consider the light emitted from the $(i,j)$th LED with the incident angle being $\varphi_{ij}$.
From \eqref{eq:varphi2},
its refractive angle with respect to the refractive center light of the $(i,j)$th LED is
\begin{align}\label{eq:psi_ij}
    \psi_{ij} = {\varphi_{ij}} \left( \frac{1}{n} + \frac{z_{S}}{\frac{R}{n-1}}- \frac{(n-1)^2 z_p}{n R}  \right).
\end{align}
Denote the ratio of the refractive angle over the incident angle by
\begin{align}\label{eq:r1}
    r = \frac{\psi_{ij} }{ {\varphi_{ij}}} =  \frac{1}{n} + \frac{z_S(n-1)}{R} - \frac{(n-1)^2 z_p}{nR} ,
\end{align}
which only depends on the vertical positions of the LED and the lens and thus is identical for all LEDs on the same XOY plane. 
Note that, in practice, the lights emitted from one LED are generally concentrated within a limited angle.
When the LED follows the perfect Lambertian distribution ($m=1$), the limited angle is $90^\circ$. When the LED is with high Lambertian order $m$, the limited angle is less than $90^\circ$. Moreover, the intensity of the lights with a large emitted angle is usually very low or may not impinge on the lens. Then, we can focus on the lights within a limited angle, which is denoted as $\Phi_{ij}$.
Hence, the intensity of the refractive light at refractive angle $\psi_{ij}$ can be expressed as
\begin{align}\label{eq:I_approx}
    I_{ij}(\psi_{ij}) &= T_{lens}(r^{-1}\psi_{ij},\psi_{ij}) I_0 \left(r^{-1}\psi_{ij}\right) U(\Phi_{ij} - r^{-1}\psi_{ij}) \nn\\
    &= T_{lens}(\varphi_{ij},\psi_{ij}) I_{0} (\varphi_{ij})U(\Phi_{ij} - \varphi_{ij}) ,
\end{align}
where $U(\cdot)$ denotes the unit step function, and
$T_{lens}(\varphi_{ij},\psi_{ij})$ is the optical lens gain which
depends on the incident angle $\varphi_{ij}$ and the refractive angle $\psi_{ij}$.
The exact value of $T_{lens}(\varphi_{ij},\psi_{ij})$ can be calculated according to \cite{6497451}.
In practice, when semi-angle of half power of an LED is small (e.g. $30^\circ$),
the optical lens gains for the lights within this angle approach a constant.
Hence, $T_{lens}(\varphi_{ij},\psi_{ij})$ can be reasonably approximated by a constant $T_{lens}$.
\begin{de}
    The {\em{beam width}} is defined as the maximal angle between the refractive lights from one LED. 
\end{de}

According to \eqref{eq:psi_ij}, the beam width of the $(i,j)$th LED is given by
\begin{align}\label{eq:beamwidth}
    \Psi_{ij} = 2 r \Phi_{ij},
\end{align}
where $\Phi_{ij}$ is the limited angle of LED $(i,j)$. 
The beam width $\Psi_{ij}$
includes all light rays emitted from the $(i,j)$th LED passing through the lens.
Other definitions of beam width exist, such as the half power beam width.
\begin{rem}
    The beam width depends on the vertical position $z_S$.
    For hemispherical lenses \eqref{eq:hemispherical},
    when $z_S$ is larger than $-R/(n(n-1))$, the ratio $r$ is between $0$ and $1$,
	implying that the beam width $\Psi_{ij}$ is smaller than 
	the width\footnote{
    {
    When $z_S$ is smaller than $-R/(n(n-1))$,
    the lights will focus to a real image of the light source,
    and thus, diverge to different directions.
    In this case, no beam can be generated.}
    } of limited angle $2\Phi_{ij}$.
    For thin lenses in \eqref{eq:thin}, when $z_S$ is larger than ${(n-2)R}/{(n-1)}$,
    the distance between the LED and the lens is less than the focal length\footnote{{
    Similarly, when $z_S$ is smaller than ${(n-2)R}/{(n-1)}$,
    the lights will focus to a real image and diverge.}
    }, i.e., $R-z_S \le f$.
    In this case, the ratio $r$ is less than $1$ too,
    and thus $\Psi_{ij}$ is also smaller than $2\Phi_{ij}$.
    Therefore, using lenses can effectively generate narrow light beams.
\end{rem}

\begin{figure}[htbp]
\centering
\begin{minipage}[t]{0.48\textwidth}
\centering
{\includegraphics[width=1\textwidth]{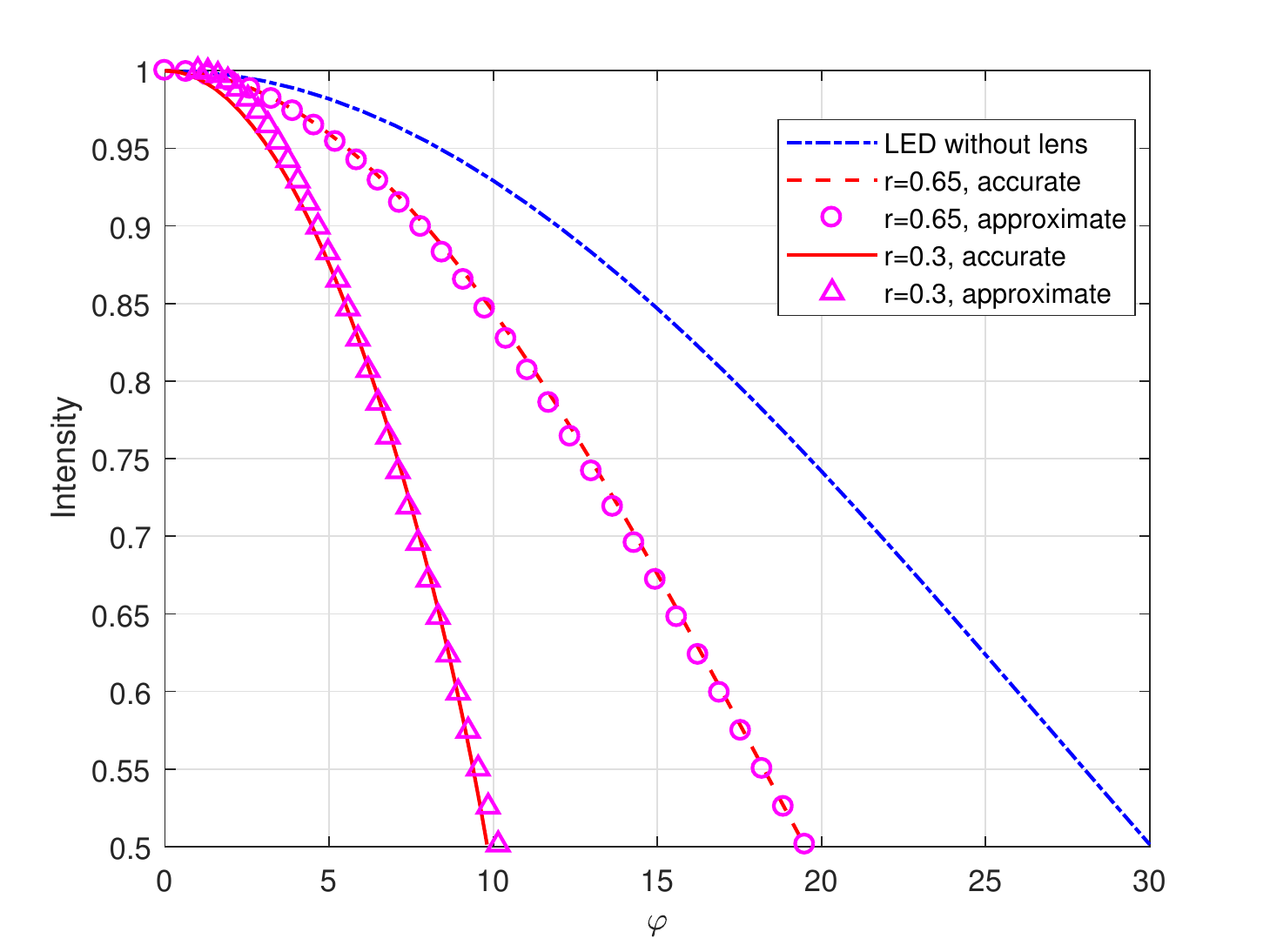}}
\caption{Comparison of the approximate and accurate light intensities.}\label{fig:Approx2}
\end{minipage}
\begin{minipage}[t]{0.48\textwidth}
\centering
{\includegraphics[width=1\textwidth]{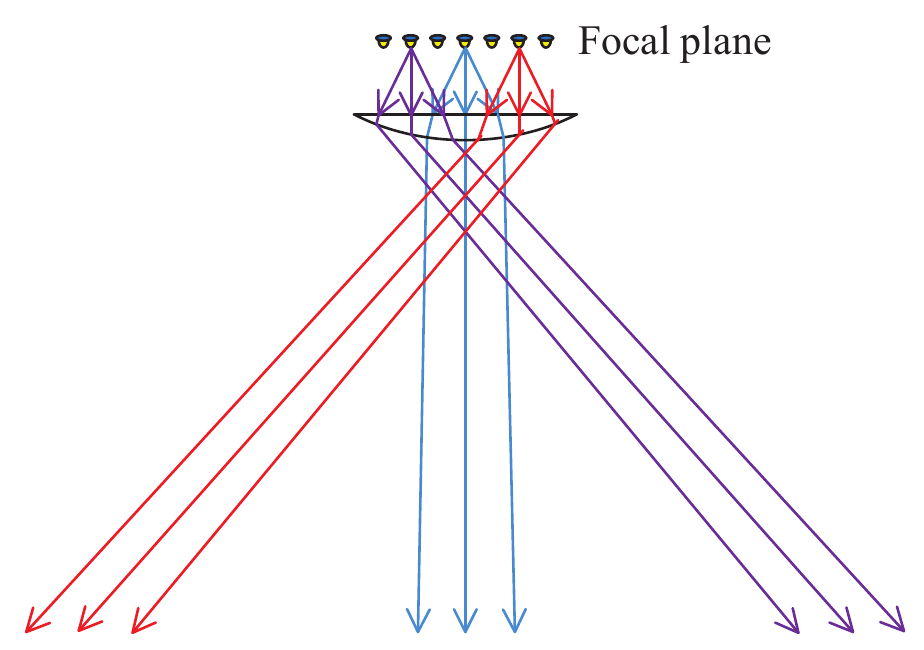}}
\caption{LEDs on the focal plane passing through the lens.}\label{fig:Parallel}
\end{minipage}
\end{figure}

In Fig.~\ref{fig:Approx2}, we compare the approximate intensity with the accurate intensity for an LED with $\varphi_C=30^\circ$.
The refractive index is $n=1.5$, and the radius of the spherical surface is $R=10$ cm.
One can observe that the approximation well approaches the accurate intensity distribution in all cases.


\subsection{Asymptotic Properties} 
\label{sub:asymptotic_properties}

Consider a square array with $M\times M$ LEDs and
suppose that the maximum illumination angle of $M$ LEDs along the x-axis is $\omega$.
Divide $\omega$ equally into $M$ small angles, each of size $\omega/M$.
To support multi-user communications, it is expected that 
each angle is covered by one beam emitted from one LED.
Then, the beam width of each LED shall be $\Psi_{ij}=\omega/M$.
With the limited angle of LED $(i,j)$ $\Phi_{ij}$
and from \eqref{eq:beamwidth},
the ratio is given by $r=\omega/(2M \Phi_{ij})$.
The $M$ LEDs along the y-axis are similar.
For a given distance $d$ between two adjacent LEDs,
the focal length of the lens shall be $f = Md/\omega$,
and according to \eqref{eq:varphi2},
the positions of LEDs can be calculated as
\begin{align}
    x_{S,i} &= \left(  -{(M-1)}/2 + (i-1)  \right)d, \quad i = 1,2,\cdots,M,  \nn\\
    y_{S,j} &= \left(  -{(M-1)}/2 + (j-1) \right)d, \quad j = 1,2,\cdots,M, \nn\\
    z_S &= \left( \frac{\omega}{2\Phi_{ij} } + \frac{\omega(n-1)z_p}{nd} - \frac{M}{n}   \right)
    \frac{d}{\omega}.
\end{align}
As the number of LEDs grows to infinity,
for hemispherical lenses, $z_p$ is $0$,
and the vertical position of LEDs tends to
\begin{align}\label{eq:21}
    \lim_{M\to\infty} z_{S} + Md/(\omega n) = 0.
\end{align}
For thin lenses, $z_p$ approaches $R$, and
the vertical position of LEDs tends to
\begin{align}
    \lim_{M\to\infty} z_{S}- (n-2)Md/\omega = 0.
\end{align}

\begin{rem}
    Since LED chips are small, a common LED array could contain a large number of LEDs,
    such as $30\times 30$ LEDs \cite{Gao2015Low} and $60\times 60$ LEDs \cite{Zeng2009High}.
    Therefore, the asymptotic property of massive LED arrays is meaningful. As $M$ increases, 
    for thin lenses, the distance between the lens and the LED array
    becomes
    $R-z_S = (n-1)Md/\omega -  (n-2)Md/\omega = f$,
    implying that the LED array is located at the focal plane of the lens.
    Hence, the lights emitted from each LED with different angles
    are refracted to one direction and become parallel, as shown in Fig.~\ref{fig:Parallel}.
    Consequently, the lights from either one LED or different LEDs are no longer interfered mutually.
\end{rem}

Next, we consider the asymptotic intensity when the number of LEDs grows to infinity.
From \eqref{eq:I_0} and \eqref{eq:I_approx},
the asymptotic intensity of a light emitted from LED $(i,j)$ with relative refractive angle $\psi_{ij}$ can be expressed as
\begin{align}\label{eq:limI}
    \lim_{M\to\infty}I_{ij}(\psi_{ij}) &= \lim_{M\to\infty} T_{lens}\frac{m_L+1}{2\pi} \cos^{m_L} \left(\frac{2M \Phi_{ij}}{\omega}\psi_{ij}\right) 
    U\left(\frac{\omega}{2M \Phi_{ij}}\Phi_{ij}- \psi_{ij}\right) \nn\\
    &= T_{lens}\frac{m_L+1}{2\pi} \delta\left(\psi_{ij}\right).
\end{align}
For this reason, the lights from LED $(i_1,j_1)$ and LED $(i_2,j_2)$ to UT $k_1$ and UT $k_2$ have the following relationship
\begin{align}
    \lim_{M\to \infty }
     I_{i_1j_1}(\psi_{i_1j_1,k_1})I_{i_1j_1}(\psi_{i_1j_1,k_2})
    =& T_{lens}^2\left( \frac{{m_L}+1}{2\pi}\right)^2 
    \delta(\psi_{i_1j_1,k_1})\delta(\psi_{i_1j_1,k_2}) = 0, \label{eq:lim_I1}\\
    \lim_{M\to \infty }I_{i_1j_1} (\psi_{i_1j_1,k_1})I_{i_2 j_2} (\psi_{i_2j_2,k_1}) =& T_{lens}^2\left( \frac{{m_L}+1}{2\pi}\right)^2
    \delta(\psi_{i_1j_1,k_1})\delta(\psi_{i_2j_2,k_1}) = 0,\label{eq:lim_I2}
\end{align}
where $\psi_{i_1j_1,k_1}$ ($\psi_{i_1j_1,k_2}$) is the angle between UT $k_1$ (UT $k_2$) and the center light of LED $(i_1,j_1)$, and $\psi_{i_2j_2,k_1}$ is the angle between UT $k_1$ and the center light of LED $(i_2,j_2)$.

\begin{rem}
    From \eqref{eq:lim_I1} and \eqref{eq:lim_I2}, we find an interesting and favorable orthogonality
    in terms of light intensities emitted from a massive LED array via a lens. 
    Specifically, for a large number of LEDs, 
    light intensities from one LED to different UTs tend to be orthogonal from \eqref{eq:lim_I1},
    while the intensities from different LEDs to one UT also tend to be orthogonal from \eqref{eq:lim_I2}.
    This is consistent with our finding in  Remark 3,
    where lights refracted by the lens tend to be parallel for a massive LED array.
    Such an asymptotic orthogonality is particularly desirable in multi-user communications, which implies that it is possible to transmit light signals to different users without interference.
\end{rem}

\subsection{Channel Gains} 
\label{sub:channel_vector}

In this subsection, we investigate channel gains of the considered optical massive MIMO system. 
We focus on the LOS propagation.
Let $A_k$ be the physical area of the photodetector of UT $k$, $d_k$ be the distance between the center of the transmit lens and the photodetector center of UT $k$, $\gamma(d_k)$ be the normalized factor to satisfy the energy conservation law of light transmission, $\psi_{ij,k}$ denote the angle between UT $k$ and the center light emitted from LED $(i,j)$, and $\phi_k$ denote the angle of incidence at UT $k$.
The channel gain between LED $(i,j)$ and UT $k$ at angles $\psi_{ij,k}$ and $\phi_k$ is
\begin{align}
    h_{k,ij} = A_k \gamma(d_k)I_{ij}(\psi_{ij,k}) \cos(\phi_k).
\end{align}
According to the energy conservation law of light transmission \cite{Ding:08},
the energy on the receiver plane is equal to the energy passing through the lens,
\begin{align}
    \int_0^{2\pi} d \zeta \int_0^{r\Phi_{ij}}  \gamma(d_k)  I_{ij} (\psi_{ij}) d_k^2\sin(\psi_{ij}) d \psi_{ij} =
    \int_0^{2\pi} d \zeta \int_0^{\Phi_{ij}} T_{lens} I_0 (\varphi_{ij}) \sin(\varphi_{ij}) d \varphi_{ij} .
\end{align}
Recalling $\psi_{ij} = r \varphi_{ij}$ and using \eqref{eq:I_approx}, the left-hand integral can be expressed as
\begin{align}
    \int_0^{r\Phi_{ij}}   I_{ij} (\psi_{ij})\sin(\psi_{ij}) d \psi_{ij} = \int_0^{\Phi_{ij}}  T_{lens} I_{0} (\varphi_{ij})\sin( r \varphi_{ij}) r d \varphi_{ij} .
\end{align}
For a large number of LEDs, $r$ is small and $\sin( r \varphi_{ij}) \approx r\sin( \varphi_{ij})$.
Therefore, we obtain $\gamma(d_k) = \frac{1}{d_k^2 r^2}$,
and the channel gain can be expressed as
\begin{align}\label{eq:h}
    h_{k,ij} = \frac{A_k}{d_k^2 r^2}  I_{ij}(\psi_{ij,k}) \cos(\phi_k).
\end{align}
Then, the channel vector $\h_k$ of UT $k$ is given by
\begin{align}\label{eq:hvec}
    \h_k =  \frac{A_k}{d_k^2 r^2}   \cos(\phi_k) 
    \begin{bmatrix}
        I_{11}(\psi_{11,k}) & I_{12}(\psi_{12,k}) & \cdots & I_{MM}(\psi_{MM,k})
    \end{bmatrix}^T.
\end{align}
Combining all user channel vectors, the multi-user channel matrix $\H \in \Cb^{K\times M^2}$ is constructed as
\begin{align}\label{eq:rank1}
    \H = \begin{bmatrix}
        \h_1 & \h_2 &  \cdots &  \h_K
    \end{bmatrix}^T.
\end{align}
\begin{rem}
    From \eqref{eq:h}, the channel gain is inversely proportional to both $d_k^2$ and $r^2$.
    When the beam width becomes narrower, the coefficient $r$ becomes smaller
    and the energy of the beam is thus concentrated in one direction.
    Moreover, according to \eqref{eq:limI}, lights from one LED become parallel,
    and different beams illuminate different directions.
    As the number of LEDs tends to infinity, only one element of the light intensities in channel vector $\h_k$
    is nonzero, implying that each UT receives signal from only one LED.
    According to \eqref{eq:lim_I1}, we have
    \begin{align}
        \lim_{M\to\infty} \h_{k_1}^T \h_{k_2} = 0,
    \end{align}
    implying that channel vectors of different UTs are asymptotically orthogonal
    in the optical massive MIMO system. Consequently, the multi-user channel matrix $\H$ has full row rank.
    In practice, the number of LEDs is generally larger than the user number;
    thus, the rank of the channel matrix $\H$ is $K$, indicating that 
    $K$ UTs can be simultaneously served by the BS.
\end{rem}
}

\section{Linear Precoding Based Transmission} 
\label{sec:transmit_design}

\subsection{MRT/RZF Linear Precoding} 
\label{sub:linear_precoding}

To communicate with a number of UTs simultaneously, BS transmits the summation of all UTs' signals,
and the signal of the $k$th UT $\x_k$ is obtained from symbols through a precoding vector $\w_k \in \Cb^{M^2\times 1}$, i.e.,
\begin{align}
	\x_k = \w_k s_k  ,
\end{align}
where $s_k$ is the real message-bearing independent and identically distributed (i.i.d.) symbols with zero-mean\footnote{As the transmit signal may take negative values,
a DC bias should be added to guarantee a non-negative input signal to the LEDs.
Assume that the LEDs transmit optical signals with a large DC bias corresponding to the working point of the LEDs.
In this way, the bipolar electrical signals are transmitted via the unipolar optical intensity.} and  unit variance.
Here, we consider two different linear precoding strategies $\w_k$ of practical interest, namely MRT $\w_k^{\rm{MRT}}$ and RZF $\w_k^{\rm{RZF}}$, which we define, respectively, as
\begin{align}
	\w_k^{\rm{MRT}} = \sqrt{\beta^{\rm{MRT}}} \h_k, \quad
	\w_k^{\rm{RZF}} =\sqrt{\beta^{\rm{RZF}}}  \left( \H^T\H + \alpha^{\rm{RZF}} \I_{N^2} \right)^{-1} \h_k  ,\label{eq:RZF}
\end{align}
where $\alpha^{\rm{RZF}} >0$ is a regularization parameter, $\beta^{\rm{MRT}}$ and $\beta^{\rm{RZF}}$ normalize the total transmit power to $\sum_k \Eb \{ \x_k^T \x_k  \}=P$, i.e.,
\begin{align}
    \beta^{\rm{MRT}} = \frac{P}{\sum_k \h_k^T\h_k},  \quad
    \beta^{\rm{RZF}} = \frac{P}{\tr \left( \H^T\H \left( \H^T\H + \alpha \I  \right)^{-2} \right) } .
\end{align}
{
The received signal-to-interference-plus-noise ratio (SINR) at UT $k$ is given by
\begin{align}
    \mbox{SINR}_k=\frac{(\h_k^T\w_k)^2}{1+\sum_{k'\neq k}(\h_k^T\w_{k'})^2}.
\end{align}
From \cite{7929273,5238736}, a lower bound and an upper bound of the achievable sum rate are given respectively by
\begin{align}
    R_{\tlb} &= \frac{1}{2}\sum_k \log(1+ \frac{6}{\pi e} \mbox{SINR}_k), \\
    R_{\tub} &= \frac{1}{2}\sum_k \log(1+\mbox{SINR}_k),
\end{align}
where the lower bound is derived from the uniform distribution,
and the upper bound is derived from the Gaussian distribution.
As the upper bound and the lower bound have the same structure,
    we introduce the following approximate sum rate expression
\begin{align}\label{eq:sum_rateR}
    R_{\tsum}= \frac{1}{2}\sum_k \log(1+ \gamma \mbox{SINR}_k),
\end{align}
where $\gamma$ is a coefficient satisfying $\frac{6}{\pi e} \le \gamma \le 1$.
\begin{rem}
    When $\gamma = \frac{6}{\pi e}$, the approximate sum rate expression represents the lower bound \cite{5238736},
    and when $\gamma=1$, the expression represents the upper bound \cite{7929273}.
    The approximate sum rate expression in \eqref{eq:sum_rateR} also coincides with the fitting function of the capacity in \cite{7397995}.
    As the expression in \eqref{eq:sum_rateR} contains the upper and lower bounds as special cases, the precoding design can maximize the upper/lower bound.
\end{rem}

Next, we consider $M$ growing infinitely large and keeping total transmit power $P$ constant.
As $M$ increases,
we can derive the approximate sum rate for MRT transmission as
\begin{align}
    \lim_{M \to \infty} \left( R^{\rm{MRT}} - \Ra^{\rm{MRT}} \right)  = 0,
\end{align}
where $\Ra^{\rm{MRT}}$ is given by
\begin{align}\label{eq:MRT}
    \Ra^{\rm{MRT}} =\frac{1}{2}\sum_k \log\left( \gamma P \frac{M^4 g_{k}^4}{\sum_{k'} g_{k'}^2}\right),
\end{align}
and from \eqref{eq:hvec}, $g_k$ is 
\begin{align}
    g_{k} = T_{lens}\frac{A_k 4\Phi^2  }{ \omega^2 d_k^2} \frac{(m_L+1) }{2\pi}  \cos(\phi_k).
\end{align}
Similarly, the approximate sum rate for RZF precoding is 
\begin{align}
    \lim_{M \to \infty} \left( R^{\rm{RZF}} - \Ra^{\rm{RZF}} \right)  = 0,
\end{align}
where
\begin{align}
    \Ra^{\rm{RZF}} =\frac{1}{2}\sum_k \log\left( \gamma\frac{PM^4}{\sum_{k'}g_{k'}^{-2} }  \right).
\end{align}

\begin{rem}
    From \eqref{eq:MRT}, the simplest MRT strategy can suppress the inter-user interference and the asymptotic sum rate increases with $M$, which is similar as RF massive MIMO systems \cite{Marzetta}.
    From the above analysis, both precoding schemes can vanish inter-user interference with an infinite number of transmit LEDs. However, due to the different power normalization factors, the sum rates of MRT and RZF are not necessarily identical. Next, we will analyze the optimal transmit signal design maximizing the sum rate.
\end{rem}

\subsection{Linear Precoding for Sum-Rate Maximization} 
\label{sub:transmit_covariance_design}

Let $\Q_k = \Eb \{ \x_k\x_k^T \}$ be the covariance matrices of transmitted signals,
and $\Q = \sum_k \Q_k$ be the covariance matrix of the sum of the transmitted signals.
{The approximate sum rate is given by}
\begin{align}\label{eq:sum1}
	R_{\tsum} = \frac{1}{2} \sum_k \left( \log\left(\Rc_k + \gamma\h_k^T  \Q_{k}  \h_k \right) - \log\left(\Rc_k\right) \right) ,
\end{align}
where $\Rc_k = 1 + \h_k^T \left( \sum_{k'\neq k} \Q_{k'} \right) \h_k$.
Let $\Qc = \{\Q_1,\Q_2,\ldots,\Q_K\}$ be the set of transmit covariance matrices.
Our main objective is to design the transmitted covariance matrices $\Qc$ maximizing the approximate sum rate $R_{\tsum}$. 
A typical power constraint is the total power constraint, which can be expressed as $\sum_k \tr(\Q_k) \le P$, where $P$ is the total power.
Define 
\begin{align}
    f(\Qc) &= \frac{1}{2}\sum_k \log \left(\Rc_k + \gamma\h_k^T  \Q_{k}  \h_k \right) ,\\
    g(\Qc) &= \frac{1}{2}\sum_k \log \left( \Rc_k\right).
\end{align}
The optimization problem under the total power constraint can be formulated as
\begin{align}\label{eq:problem1}
	\max_{\Qc} \quad & f(\Qc) - g(\Qc)\nn                  \\
	\tst \quad        & \sum_k \tr\left(\Q_k  \right) \le P, \quad 
                        \Q_k \succeq \bb.       
\end{align} 
Due to the concavity of $\log(\cdot)$ function, the approximate sum rate $R_{\tsum}$ is a difference of concave functions (d.c.).
To solve problem \eqref{eq:problem1}, we utilize the concave-convex-procedure (CCCP), which is an iterative procedure solving a sequence of convex programs.
The idea of CCCP program is to linearize the concave part around a solution obtained in the current iteration.
Employing the CCCP method, the iterative procedure is expressed as
\begin{align}\label{eq:CCCP1}
    \left[ \Qc^{(i+1)}\right] = & \arg\max_{\Qc} f(\Qc) - \sum_k \tr \left( \left( \frac{\partial}{\partial \Q_k} g\Big(\Qc^{(i)}\Big)\right)^T \Q_k \right)  \nn\\
    \tst \quad & \sum_k \tr \left( \Q_k \right) \le P, \quad
    \Q_k \succeq \bb.
\end{align}
To understand the properties of this convex optimization problem better, we present some
basic properties of the generated sequences by \eqref{eq:CCCP1}.
\begin{thm}\label{thm:CCCP1}
    Let $\left\{ \Qc^{(i)} \right \}_{i=0}^{\infty}$ be any sequences generated by \eqref{eq:CCCP1}.
    Then, all limit points of  $\left\{ \Qc^{(i)} \right \}_{i=0}^{\infty}$ are stationary points of the d.c. program in \eqref{eq:problem1}.
    In addition, $\lim_{i\to\infty} \left( f(\Qc^{(i)}) - g(\Qc^{(i)} \right)$ $ = f(\Qc^{(*)}) - g(\Qc^{(*)})$,
    where $\left\{\Qc^{(*)}\right\}$ is a stationary point of problem \eqref{eq:problem1},
    {and the rank of $\Q_{k}^{(*)}$ satisfies $\rank(\Q_k^{(*)})\le 1$.}
    \begin{IEEEproof}
        See Appendix \ref{sec:proof_of_theorem_thm:cccp1}.
    \end{IEEEproof}
\end{thm}

In optical communications, the common IM/DD schemes require driving current must be non-negative.
Thus, the transmit current of each LED is limited to guarantee the non-negative input signal. This constraint can be expressed as
\begin{align}\label{eq:w_constraint}
	\sum_k [\w_k]_m \le b, \quad m= 1,2,\ldots,M^2,
\end{align}
where $b$ is the bias current. Employing the result in \cite{7247365}, we have
\begin{align}
	\frac{\left( \sum_k [\w_k]_m \right)^2}{K} \le \sum_k [\w_k\w_k^T]_{mm} = \sum_k \left[ \Q_k \right]_{mm}.
\end{align}
Thus, $\sum_k \left[ \Q_k \right]_{mm} \le {b^2}/{K}$
 can ensure the constraint in \eqref{eq:w_constraint}.
Here, we consider the transmit design problem under the per LED power constraint.
Define unit vector $\e_m = \left[ 0,\ldots,0,1,0,\ldots,0 \right]^T$,
where only the $m$th element is 1. With the definition of $f(\Qc)$ and $g(\Qc)$, 
the transmit design problem under the per LED power constraint can be expressed as
\begin{align}\label{eq:op2}
	\max_{\Qc} \quad &  f(\Qc) - g(\Qc) \nn                  \\
	\tst \quad        & \Q_k \succeq \bb,\quad \e_m^T \left( \sum_k \Q_k \right) \e_m \le p,  \quad m=1,2,\cdots,M^2  .     
\end{align}
where $p={b^2}/{K}$ is the maximal power per LED.
Similar to problem \eqref{eq:problem1}, due to the concavity of $f(\Qc)$ and $g(\Qc)$ on $\Q_k$,
problem \eqref{eq:op2} is a d.c. program.
Utilizing the CCCP method, we can solve the d.c. problem by iteratively solving the following convex problem:
\begin{align}\label{eq:CCCP2}
	\left[ \Qc^{(i+1)} \right] = & \arg\max_{\Qc} f(\Qc)   - \sum_k \tr \left( \left( \frac{\partial}{\partial \Q_k} g\Big(\Qc^{(i)}\Big)\right)^T \Q_k \right)  \nn \\
	\tst \quad & \Q_k \succeq \bb,\quad \e_m^T \left( \sum_k \Q_k \right) \e_m \le p,  \quad m=1,2,\cdots,M^2  .  	
\end{align}
For the generalized sequences by iteratively solving problem \eqref{eq:CCCP2},
we have the following result.
\begin{thm}\label{thm:CCCP2}
    Let $\left\{ \Qc^{(i)} \right \}_{i=0}^{\infty}$ be any sequences generated by \eqref{eq:CCCP2}.
    Then, all limit points of  $\left\{ \Qc^{(i)} \right \}_{i=0}^{\infty}$ are stationary points of the d.c. program in \eqref{eq:op2}.
    In addition, $\lim_{i\to\infty} \left( f(\Qc^{(i)}) - g(\Qc^{(i)} \right)$ $ = f(\Qc^{(*)}) - g(\Qc^{(*)})$
    where $\left\{\Qc^{(*)}\right\}$ is a stationary point of problem \eqref{eq:op2},
    {and the rank of $\Q_{k}^{(*)}$ satisfies $\rank(\Q_k^{(*)})\le 1$.}
	\begin{IEEEproof}
		The proof is similar to that of Theorem  \ref{thm:CCCP1}, and is omitted here.
	\end{IEEEproof}
\end{thm}

\begin{rem}
    Theorem \ref{thm:CCCP1} and Theorem \ref{thm:CCCP2} indicate that iterative procedures \eqref{eq:CCCP1} and \eqref{eq:CCCP2} can find candidate optimal solutions under both power constraints.
    The solutions converge to some stationary points of the original d.c. programs.
	In \cite{Chen17}, it was shown that the CCCP algorithm
	can converge in a few iterations.
    For optical massive MIMO communications, as the number of transmit LEDs increases, the dimension of transmit covariance matrix $\Q_k$ becomes large,
    and thus, the computational complexity of each iteration is also demanding.
    In Section \ref{sec:bdma_transmission} and Section\ref{sec:bdma_transmission_under_per_led_power_constraint}, we will analyze the asymptotic performance and propose BDMA transmissions.
\end{rem}

\section{BDMA Transmission under Total Power Constraint} 
\label{sec:bdma_transmission}

In this section, we study the transmitted covariance matrix $\Q_k$ design under the total power constraint.
We present the asymptotically optimality of BDMA transmission with infinite LEDs and prove the beam orthogonality with limited LEDs.

\subsection{Asymptotic Analysis} 
\label{sub:asymptotic_analysis}


Let $\R_k = \h_k \h_k^T$.
The approximate sum rate in \eqref{eq:sum1} can be rewritten as
\begin{align}\label{eq:sum rate}
	R_{\tsum} = \frac{1}{2}\sum_k \log \left( 1 + \gamma\frac{\tr(\R_k\Q_k)}{1 + \tr \left( \R_k \sum_{k'\neq k} \Q_{k'} \right)  } \right) .
\end{align}
We can derive the asymptotic result of the sum rate $R_{\tsum}$ as follows.
\begin{thm}\label{thm:gamma}
    As $M$ goes to infinity, the sum rate $R_{\tsum}$ tends to $\Ra_{\tsum}$, i.e.,
    \begin{align}
        \lim_{M \to \infty}R_{\tsum} - \Ra_{\tsum} \to 0,
    \end{align} 
    where $\Ra_{\tsum}$ is given by
    \begin{align}\label{eq:asymp_sum_rate}
    \Ra_{\tsum} =  \frac{1}{2}\sum_k \log \left( 1 + \gamma\frac{M^4g_k^2 [\Q_k]_{m_km_k}}{1+M^4 g_k^2\sum_{k'\neq k} [\Q_{k'}]_{m_km_k}} \right),
    \end{align}
    where $m_k = (i_k-1)M+j_k$, and $(i_k,j_k)$ satisfies $\psi_{i_kj_k,k} = 0$.
\end{thm}
\begin{IEEEproof}
    See Appendix \ref{sec:proof_of_proposition_pro:1}.
\end{IEEEproof}
\begin{rem}
    Theorem \ref{thm:gamma} presents that the sum rate $R_{\tsum}$ tends to asymptotic sum rate $\Ra_{\tsum}$ when $M$ goes to infinity.
    For a large but finite $M$, $\Ra_{\tsum}$ is an approximation of the sum rate $R_{\tsum}$.
    Moreover, from \eqref{eq:asymp_sum_rate},
    the asymptotic $\Ra_{\tsum}$ only depends on the diagonal elements of $\Q_k$.
    As $M$ tends to infinity, UT $k$ receives signals from only LED $(i_k,j_k)$ ($\psi_{i_kj_k,k} = 0$),
    which is coincident with Remark 5.
    This means that for a large number of LEDs, only diagonal elements of $\Q_k$ dominant the sum rate.
\end{rem}

Then, we consider the transmit design maximizing the asymptotic sum rate $\Ra_{\tsum}$, which is given by
\begin{align}\label{eq:Prob5_1}
    \max_{\Q_1,\Q_2,\cdots,\Q_K}  &  \Ra_{\tsum}\nn\\
    \tst \quad &  \sum_k \tr \left( \Q_k \right) \le P, \quad
                \Q_k \succeq \bb.
\end{align}
As UTs are distributed in the different positions, and thus, different UTs are illuminated by different LEDs, i.e., for $k_1 \neq k_2$,
we have $(i_{k_1},j_{k_1}) \neq (i_{k_2},j_{k_2})$ and $m_{k_1} \neq m_{k_2}$.
Thus, we can have the solution of problem \eqref{eq:Prob5_1} as in the following theorem.
\begin{thm}\label{thm:4}
    The optimal covariance matrix $\Q_k$ is a diagonal matrix,
    and the diagonal elements are the water-filling solution as
    \begin{align}\label{eq:waterfill}
        \left[ \Q_k \right]_{mm} = 
        \left\{ \begin{array}{*{20}{c}}
            \left(  \frac{1}{\nu} - \frac{1}{\gamma M^4 g_k^2} \right)^+, & m =m_k, \\
              0, & m \neq m_k.     
        \end{array}
        \right.
    \end{align}
    where $(x)^+=\max \{x,0\}$, $\nu$ is the Lagrange multiplier satisfying the condition
\begin{align}
    \sum_m \left(  \frac{1}{\nu} - \frac{1}{\gamma M^4 g_k^2} \right)^+ = {P} .  
\end{align}
In the limit of large $M$, the optimal sum rate $R_{\tsum}^{o}$ can be expressed as
\begin{align}\label{eq:asymptotic_sum_rate}
    \lim_{M\to \infty}R_{\tsum}^{o} - \frac{1}{2} \sum_k\log \left( 1 +  \gamma M^4 g_k^2  [\Q_k]_{m_km_k}  \right) = 0 .
\end{align}
\end{thm}
\begin{IEEEproof}
    See Appendix \ref{proof:thm4}.
\end{IEEEproof}
\begin{rem}
    For a large $M$, the solution \eqref{eq:waterfill} can maximize the sum rate $R_{\tsum}$, which is asymptotically optimal.
    The asymptotically optimal transmit covariance matrix $\Q_k$ should be a diagonal matrix. This means that beams generated by different LEDs transmit independent signals, which is called beam domain transmission. Moreover, in the beam domain transmission, different beams serve different UTs and beams for different UTs are non-overlapping, which is called BDMA transmission\cite{BDMA,You17BDMA}.
    These results show that BDMA transmission is asymptotically optimal under the total power constraint. 
    Theorem \ref{thm:4} also shows that with a large number of LEDs, the performance of the MU-MISO system is asymptotically equal to the summation performance of $K$ SU-SISO systems, without any inter-user interference.
\end{rem}

\subsection{Comparison with the Case without Transmit Lens} 
\label{sub:comparison_of_led_array_without_transmit_lens}

The channel vector without a transmit lens $\hht_k$ can be expressed as
\begin{align}
    \hht_k = \frac{A_k}{d_k^2}I_0(\varphi_k) \cos(\phi_k) {\bf 1}_{M^2\times 1} \defeq \gt_k{\bf 1}_{M^2\times 1} .
\end{align}
where $\phi_k$ is the angle of incidence at UT $k$.
Let $\Rt_k = \hht_k\hht_k^T = \gt_k^2 {\bf {1}}_{M^2\times M^2}$ .
Then, the transmit design problem is given by
\begin{align}\label{eq:compare1}
    \max_{\Q_1,\Q_2,\cdots,\Q_K} \quad &  \frac{1}{2}\sum_k \left(\log\left(1 + \tr( \Rt_k\Q)-(1-\gamma) \tr(\Rt_k\Q_k)  \right) - \log\left(1 +\tr( \Rt_k \left( \Q - \Q_k \right) )\right)\right)  \nn \\
    \tst \quad        & \sum_k\tr\left( \Q_k  \right)  \le P ,  \quad
                       \Q_k \succeq \bb.  
\end{align}
The optimal solution can be obtained as
\begin{align}
    \Q_k= \left\{  \begin{array}{*{20}{c}}
        \frac{P}{M^2} \I_{M^2}, & k = \arg\max_{k'}\gt_{k'}^2, \\
        \bb, & k \neq  \arg\max_{k'}\gt_{k'}^2 .
    \end{array}
    \right.  
\end{align}
Thus, the sum rate of the conventional transmission without a transmit lens is
\begin{align}
    \tilde R_{\tsum} =  \frac{1}{2}\log\left(1 + \gamma M^2 \gt_k^2 P \right).
\end{align}

Now we can compare the optimal sum rate performances of transmission schemes with and without a transmit lens for the asymptotic case. 
As $M$ increases to infinity, we obtain the sum rate ratio as
\begin{align}\label{eq:Ratio1}
    \lim_{M\to \infty} \frac{R_{\tsum}^o}{\tilde R_{\tsum}} 
    &= \sum_k \lim_{M\to \infty}  \frac{ \frac{1}{2}\log \left( 1 + \gamma M^4 g_k^2  [\Q_k]_{m_km_k}  \right)}{\frac{1}{2}\log\left(1 + \gamma M^2 \gt_k^2 P \right)}  = 2K.
\end{align}
The above analysis indicates that
in the asymptotic case ($M\to \infty$), the sum rate of BDMA transmission is $2K$ times more than that of the conventional transmission without a lens.

\subsection{BDMA for Non-Asymptotic Case} 
\label{sub:bdma_for_non_asymptotic_case}

Motivated by the asymptotically optimality of BDMA transmission, 
in the beam domain, different beams transmit independent signals
and different beams transit signals to different UTs.
For non-asymptotic case, we remain focused on the beam domain transmission 
(i.e., different beams transmit independent signals).
Thus, the transmit covariance matrix $\Q_k$ design problem is degraded to a diagonal power allocation matrix $\Lambdam_k$ ($\Lambdam_k$ is a diagonal matrix with eigenvalues of $\Q_k$) optimization, which can be expressed as
\begin{align}\label{eq:natp1}
    \max_{\Lambdam_1, \Lambdam_2, \cdots, \Lambdam_K} \quad &\frac{1}{2} \sum_k \left( \log\left(1 + \tr(\R_k \Lambdam) - (1- \gamma ) \tr (\R_k \Lambdam_k) \right) - \log\left(1 + \tr\left( \R_k \left( \Lambdam- \Lambdam_{k} \right) \right) \right) \right) \nn \\
    \tst \quad              & \tr\left( \Lambdam \right)  \le P ,\quad
                            \Lambdam_k \succeq \bb,
\end{align}
where $\Lambdam = \sum_k \Lambdam_k$.
For problem \eqref{eq:natp1}, we can derive orthogonality conditions of optimal power allocation as follows.
\begin{thm}\label{thm:5}
    The optimal power allocation for each UT under the total power constraint should be non-overlapping (orthogonal) across beams,
    i.e., the solution of problem \eqref{eq:natp1} satisfies the following conditions: 
    \begin{align}\label{eq:orthogonal_conditions}
        \Lambdam_{k_1} \Lambdam_{k_2} = \bb, \quad k_1\neq k_2.
    \end{align}
    \begin{IEEEproof}
        See Appendix \ref{proof:thm5}.
    \end{IEEEproof}
\end{thm}
\begin{rem}
    With a large but limited number of transmit LEDs, if BS transmits independent signals in the beam domain,
    the optimal power allocation should be orthogonal between UTs.
    This means that one transmit beam only communicates with one UT, and transmit beams for different UTs should be non-overlapping.
    Thus, BDMA transmission is optimal for sum rate maximization in the beam domain,
    which coincides with the previous results for massive MIMO RF communications \cite{Chen17}.
\end{rem}

Next, we propose a simple beam allocation algorithm which satisfies the orthogonality conditions \eqref{eq:orthogonal_conditions}.
Consider equal power allocation for the selected transmit beams.
Define $\Lambdam_k = \eta \B_k$, where $\B_k$ is the beam allocation matrix with either $0$ or $1$ on the diagonal and $\eta$ is an auxiliary variable to satisfy the power constraint.
Assume that the maximal number of beams for each UT is $B$.
The power allocation optimization problem can be degraded to a beam allocation algorithm as
\begin{align}\label{eq:beam_allocation}
    \max_{\eta,\B_1, \B_2, \cdots, \B_K} \quad & \frac{1}{2}\sum_k \left( \log\left(1 + \tr\left(\eta\R_k \left( \sum_{k'\neq k} \B_{k'} \right)  + \gamma \tr (\eta \R_k \B_k) \right) \right) \right. \nn\\
    & \left. - \log\left(1 + \tr\left( \eta \R_k \left(\sum_{k'\neq k} \B_{k'}  \right) \right) \right)\right)  \nn \\
    \tst \quad              & \eta \sum_k \tr (\B_k) = P, \quad \tr(\B_k) \le B,  		\quad \B_k(\I-\B_k) = \bb.
\end{align}
Then, we propose a beam allocation algorithm, including the following stpdf:
\begin{enumerate}
    \item Initialize $i = 1$, $R = 0$, $\B_k = \bb$.
    \item Initialize $j = 1$, and $\d = [d_1,d_2,\cdots,d_{M^2}]$ is the index of sorted diagonal elements of $\R_i$.
    \item Set $[\B_i]_{d_{j}d_{j}} = 1$, calculate $\eta$ according to $\eta \sum_k \tr (\B_k) = P$ and $R_{\tsum}$ according to \eqref{eq:beam_allocation}.
    \item If $R_{\tsum}>R$, set $R = R_{\tsum}$, $j = j+1$, and if $j\le B$, return to Step 3; else, set $[\B_i]_{d_{j}d_{j}} = 0$ and recalculate $\eta$. 
    \item Set $i = i+1$. If $i\le K$, return to Step 2; else, stop the algorithm.
\end{enumerate}

For the beam allocation algorithm, the complexity of the algorithm is $O(B K)$. Usually, the number of beams for one UT $B$ is much smaller than the total number of LEDs $M^2$. Thus, this algorithm has a low computational complexity.

\section{BDMA Transmission under Per LED Power Constraint} 
\label{sec:bdma_transmission_under_per_led_power_constraint}

We have analyzed the optimality of BDMA transmission under the total power constraint .
In this section, we design the transmit covariance matrix $\Q_k$ under the per LED power constraint. 

\subsection{Asymptotic Analysis} 
\label{sub:asymptotic_transmit_design_with_per_led_power_constraint}

From Theorem \ref{thm:gamma}, as $M$ tends to infinity, 
the sum rate $R_{\tsum}$ tends to asymptotic sum rate $\Ra_{\tsum}$.
Then, we consider the transmit design maximizing the sum rate $\Ra_{\tsum}$ under the per LED power constraint, which can be expressed as
\begin{align}\label{eq:prob_per}
    \max_{\Q_1,\Q_2,\cdots,\Q_K}  & \Ra_{\tsum} \nn\\
    \tst \quad & \Q_k \succeq \bb , \quad \e_m^T\Q_k \e_m  \le { p}   ,  \quad m=1,2,\ldots,M^2.
\end{align}
Similarly, we can have the following asymptotically optimal transmit covariance matrices.
\begin{thm}
    The optimal transmit covariance matrix $\Q_k$ is a diagonal matrix,
    and the diagonal elements can be expressed as
    \begin{align}
        [\Q_k]_{m m}=
        \left\{ \begin{array}{*{20}{c}}
        { p}, & m =m_k, \\
        0, & m\neq m_k.
        \end{array}
        \right. 
    \end{align}
    Thus, in the limit of large $M$, the optimal sum rate $R_{\tsum}^{o}$  can be expressed as
    \begin{align}\label{eq:asymptotic_sum_rate_2}
        \lim_{M \to \infty}R_{\tsum}^{o} - \frac{1}{2}\sum_k \log \left( 1 + \gamma M^4 g_k^2  p \right) = 0 .
    \end{align}
\end{thm}

\begin{rem}
    With asymptotically large LEDs, BDMA transmission can achieve the optimal performance under the per LED power constraint.
    The asymptotic sum rate is the summation rate of $K$ SU-SISO systems without inter-user interference.
\end{rem}

\subsection{ Comparison with the Case without Transmit Lens} 
\label{sub:_comparison_with_led_array_without_transmit_lens}

To compare the performance of BDMA with the conventional transmission without a transmit lens,
we first calculate the maximal sum rate under the per LED power constraint.
Under the per LED power constraint, the optimal solution $\Q_k$ can be obtained as 
\begin{align}
    \Q_k = \left\{ \begin{array}{*{20}{c}}
    p{\bf 1}_{M^2 \times M^2}, & k = \arg\max_{k'}  \gt_{k'} ^2, \\
    0, & k \neq \arg\max_{k'} \gt_{k'}^2 .
    \end{array}
    \right. 
\end{align}
and, the optimal sum rate of the conventional transmission without a transmit lens is given by
\begin{align}
     \tilde R_{\tsum} =\frac{1}{2} \log\left(1 + \gamma M^4 \gt_k^2 p  \right).
\end{align}

Now, we can compare the performances of transmission schemes with and without a transmit lens and have
\begin{align}\label{eq:Ratio2}
    \lim_{M\to \infty} \frac{R_{\tsum}^{o}}{ \tilde R_{\tsum}}
    & = \sum_k  \lim_{M\to \infty}  \frac{\frac{1}{2}\log \left( 1 + \gamma M^4 g_k^2  p \right)}{\frac{1}{2}\log\left(1 + \gamma M^4 \gt_k^2 p  \right)}
     = K.
\end{align}
When the number of transmit LEDs goes to infinity, the sum rate of our proposed BDMA transmission is $K$ times more than that of conventional transmissions without a transmit lens.

\subsection{BDMA for Non-Asymptotic Case} 
\label{sub:non_asymptotic_transmit_design_with_per_led_power_constraint}

Motivated by the asymptotic result, we consider the beam domain transmission, where each beam transmits independent signals and the transmit design problem is degraded to a power allocation problem, which can be expressed as
\begin{align}\label{eq:natp2}
	\max_{\Lambdam_1,\Lambdam_2,\ldots,\Lambdam_K} \quad &\frac{1}{2} \sum_k \left(\log\left(\I + \tr(\R_k \Lambdam ) -(1- \gamma) \tr(\R_k \Lambdam_k)  \right) - \log\left(\I + \tr(\R_k \left( \Lambdam- \Lambdam_{k} \right) ) \right)\right)  \nn \\
	\tst \quad              & \Lambdam_k \succeq \bb, \quad \e_m^T \Lambdam \e_m \le p, \quad m = 1,2,\ldots, M^2 .
\end{align}
For the power allocation problem, we can derive the following result.
\begin{thm}\label{thm:6}
	The optimal power allocation for each UT under the per LED power constraint should be non-overlapping (orthogonal) across beams,
	i.e., the solution of problem \eqref{eq:natp2} satisfies the following conditions:
	\begin{align}\label{eq:ortho_conditions2}
		\Lambdam_{k_1} \Lambdam_{k_2} = \bb, \quad k_1\neq k_2.
	\end{align}
	\begin{IEEEproof}
		The proof is similar with that of Theorem \ref{thm:5}, and is omitted here.
	\end{IEEEproof}
\end{thm}
\begin{rem}
	For beam domain transmission, it is optimal that different LEDs transmit signals to different UTs.
	Combined with Theorem \ref{thm:5}, under both power constraints, BDMA transmission can achieve optimal performance.
\end{rem}

Consider beam allocation with equal power on the selected beams. Define $\Lambdam_k = \eta \B_k$, and the beam allocation algorithm is similar with the algorithm in Section \ref{sec:bdma_transmission}. The difference is the power factor $\eta$.
Here, we set $\eta = p$ under the per LED power constraint.

\section{Simulation Results} 
\label{sec:simulation_results}

In this section, we give some examples to illustrate the performance of our proposed optical massive MIMO communication approaches with a transmit lens, in comparison with the conventional transmission without a lens.
We consider two typical optical massive MIMO communication scenarios. One is for optical communication in a small area, such as a meeting room, where BS equipped with $12\times 12$ LEDs serves 20 UTs. The room size is $5$ m $\times$ $5$ m, and the height is $3$ m. 
{BS is located at the center of the room, and}
UTs are randomly distributed in the room.
The other is for a wide area, for example airport lounge or stadium, where BS with $80 \times 80$ LEDs serves $484$ UTs.
The area size is $16$ m $\times$ $16$ m, and the height is $8$ m.
We consider two user distributions in the wide area: randomly distributed and uniformly placed.
For the uniformly placed case, the $(i,j)$th UT position is given by
\begin{align}
    (X_i,Y_j) = (-7.6+0.69(i-1),-7.6+0.69(j-1)),\quad i,j=1,2,\cdots,22,
\end{align}
{and for the randomly distributed case, we average the results over 10000 realizations.}
We define the transmit signal-to-noise ratio (SNR) as ${\rm{SNR} }= P/\sigma^2$,
and consider the same total transmit power under both power constraint, i.e., $P=pM^2$.
A DC bias is added to guarantee a non-negative input signal.
For the uniform input distribution, the electrical power of DC bias
is $P_{DC} = 3P$. In the simulation, we consider the input of uniform distribution and calculate the sum rate lower bound.

\begin{figure}[!h]
\centering
\subfigure[]{\includegraphics[width=0.33\textwidth]{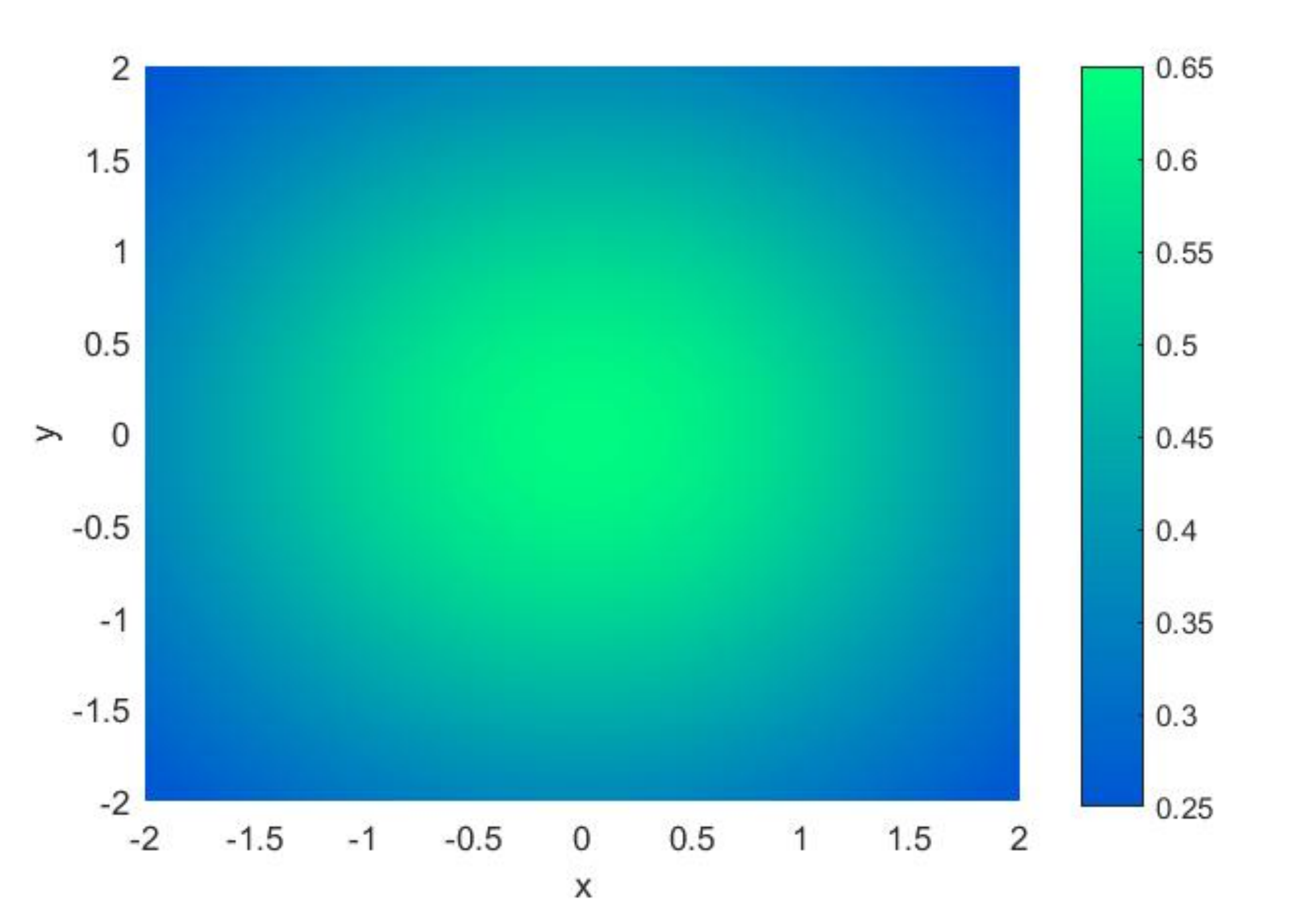}\label{fig:Channel:1}}
\subfigure[]{\includegraphics[width=0.3\textwidth]{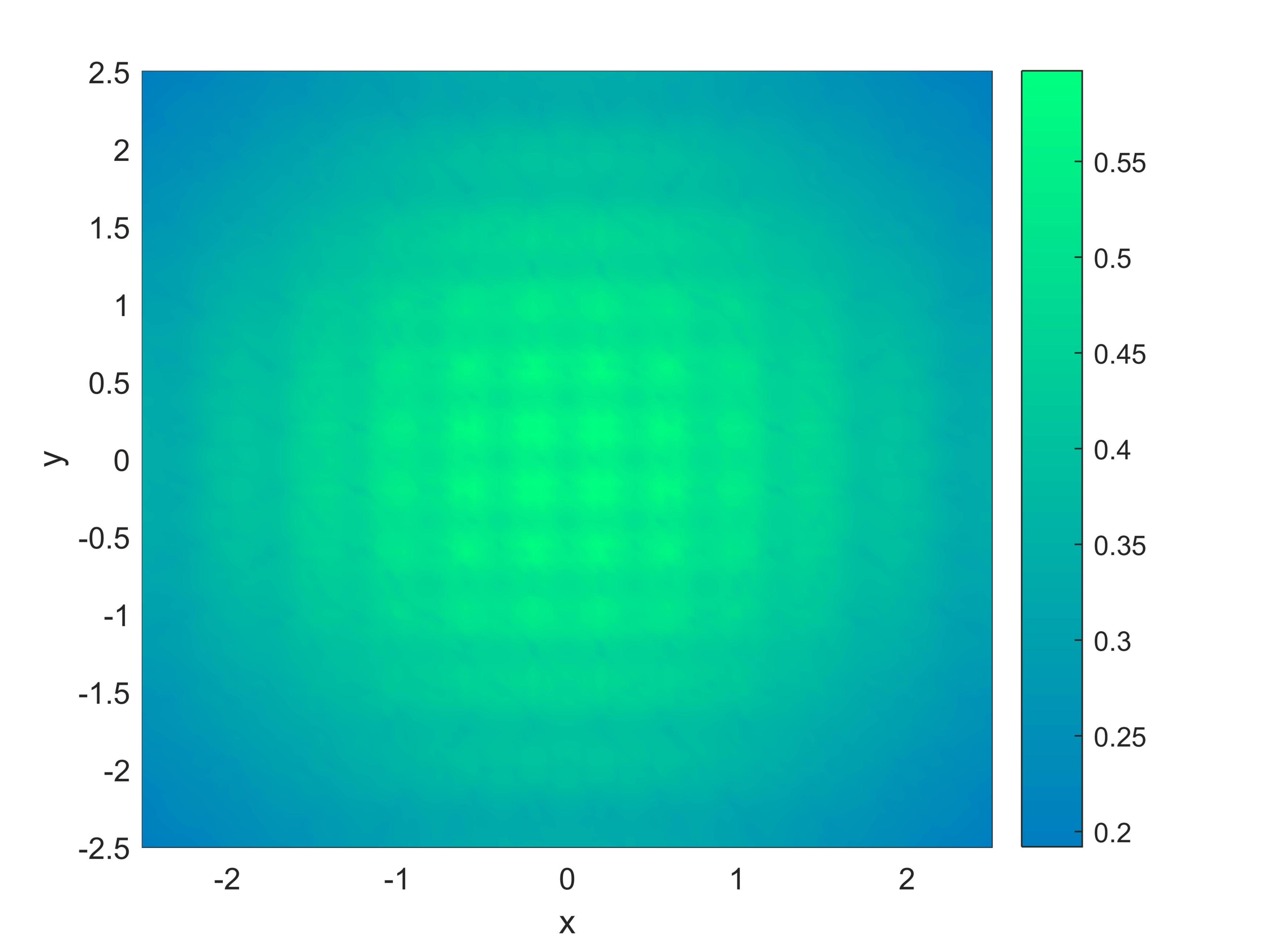}\label{fig:Channel:2}}
\subfigure[]{\includegraphics[width=0.3\textwidth]{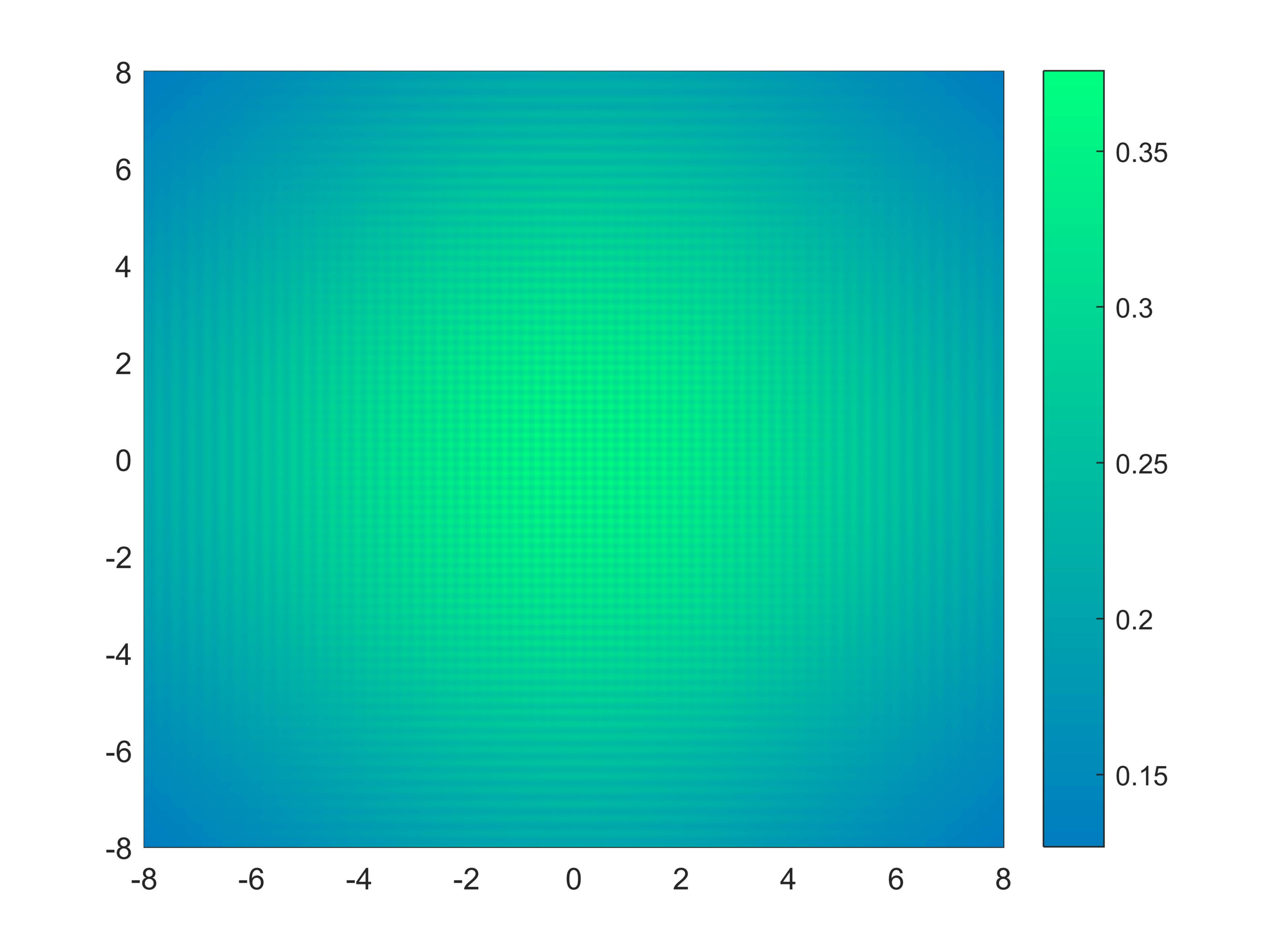}\label{fig:Channel:3}}
\caption{Optical intensity on the receiver plane, (a) $12\times 12$ LEDs without a lens, (b) $12\times 12$ LEDs with a transmit lens, (c) $80\times 80$ LEDs with a transmit lens.}\label{fig:Channel}
\end{figure}

{
Fig.~\ref{fig:Channel} compares the optical intensity on the receiver plane with and without a transmit lens,
where the LEDs transmit signals with the same power.
The optical intensity is obtained for each sample by $I_r = \frac{1}{M}{\bf 1}_{1\times M^2} \h$. 
Fig.~\ref{fig:Channel:1} and Fig.~\ref{fig:Channel:2},
respectively, illustrate the optical intensity without a lens and with a transmit lens in the small area scenario,
while Fig.~\ref{fig:Channel:3} illustrates the optical intensity with a transmit lens in the wide area scenario.
From the results, 
optical intensity distributions with and without a transmit lens are similar.
In fact, as the number of LEDs increases, the optical intensity distributions with a transmit lens tend to be more uniform.
}

\begin{figure}[htbp]
\centering
\subfigure[]{\includegraphics[width=0.45\textwidth]{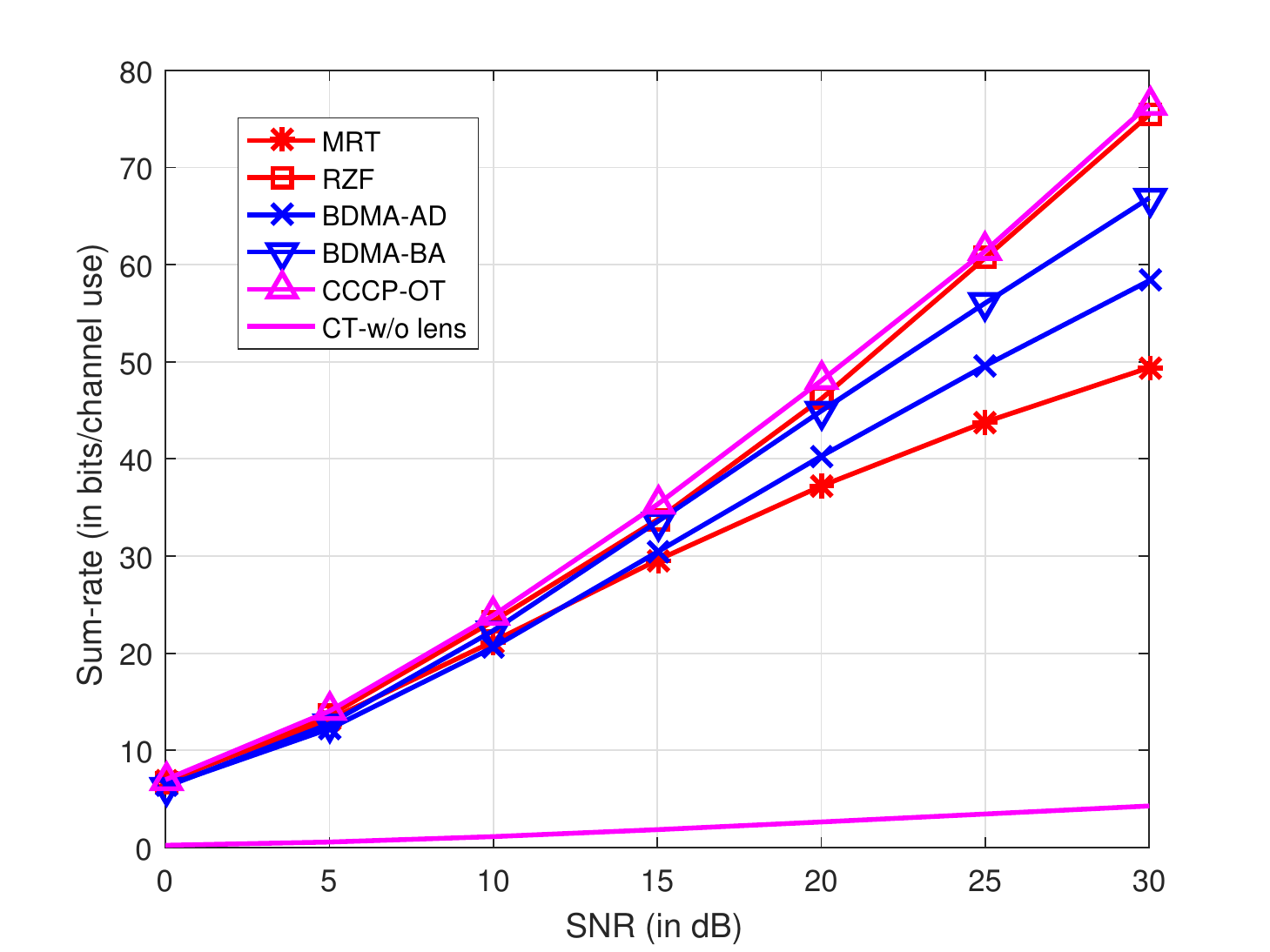}\label{fig:total}}
\subfigure[]{\includegraphics[width=0.45\textwidth]{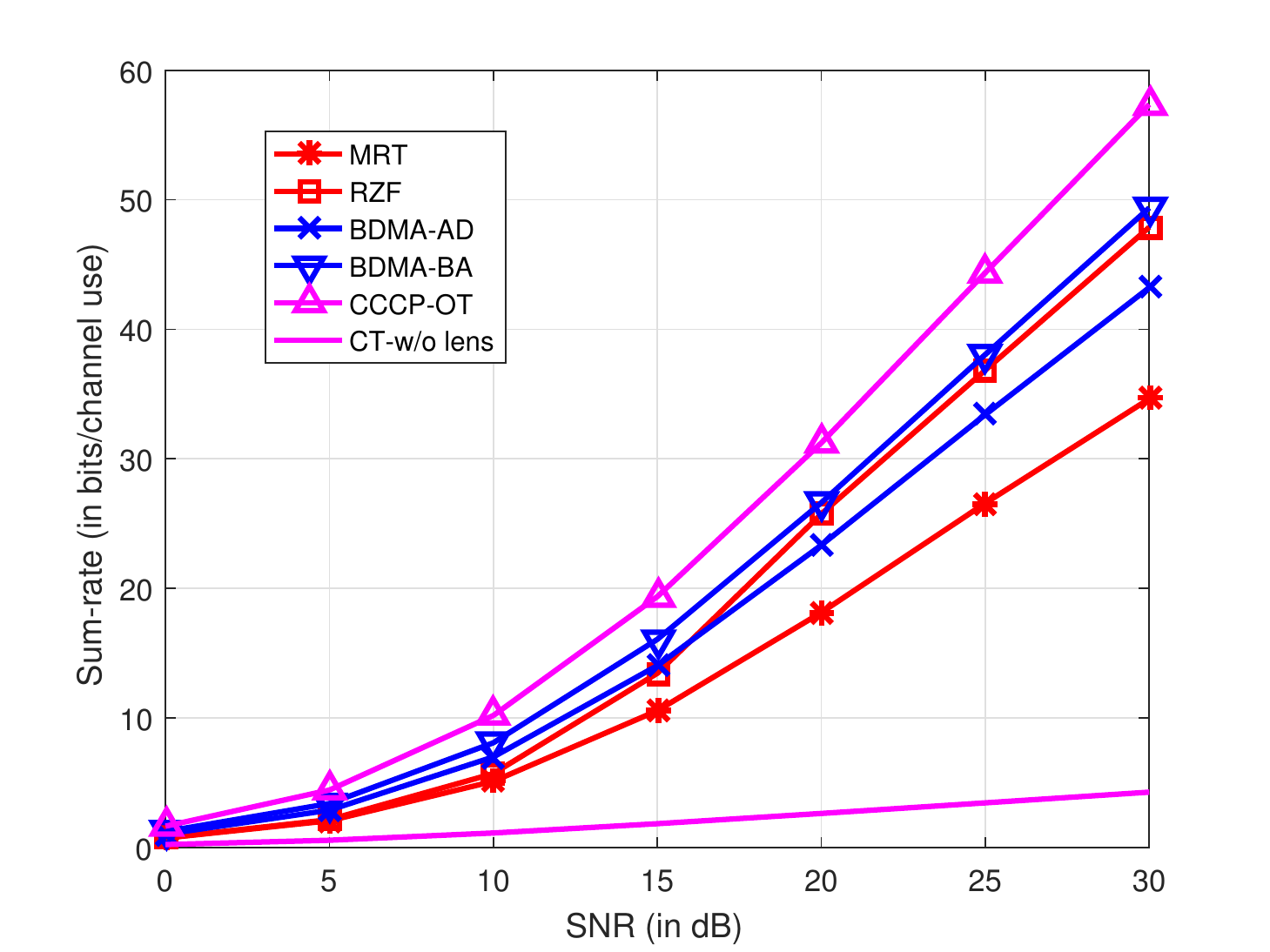}\label{fig:per}}
\caption{Comparison of sum rate in the small area scenario, (a) under the total power constraint, (b) under the per LED power constraint.}
\label{fig:4}
\end{figure}

{
Fig.~\ref{fig:4} compares the sum rates of different schemes in the small area scenario. 
The RZF precoding and the CCCP based optimized transmission (CCCP-OT) have similar performance, while
BDMA using beam allocation algorithm (BDMA-BA) can approach them with about $2.5$ dB performance loss at high SNR.
The BDMA based on asymptotic design (BDMA-AD) as in (55) has a slight performance loss compared to BDMA-BA.
In the high SNR regime, the average rate per UT approaches $4$ bits per channel use.
Under the per LED power constraint, we multiply the precoding vectors of MRT and RZF
by a power factor to make the maximal power on one LED equal to the power constraint. 
Thus, the sum rate of RZF is lower than BDMA-BA,
and the average rate per UT is about $2.5$ bits per channel use.
Under both power constraints, the sum rate of the conventional transmission without a lens (CT-w/o lens) is much smaller than that of BDMA-BA.
}

\begin{figure}[htbp]
\centering
\subfigure[]{\includegraphics[width=0.45\textwidth]{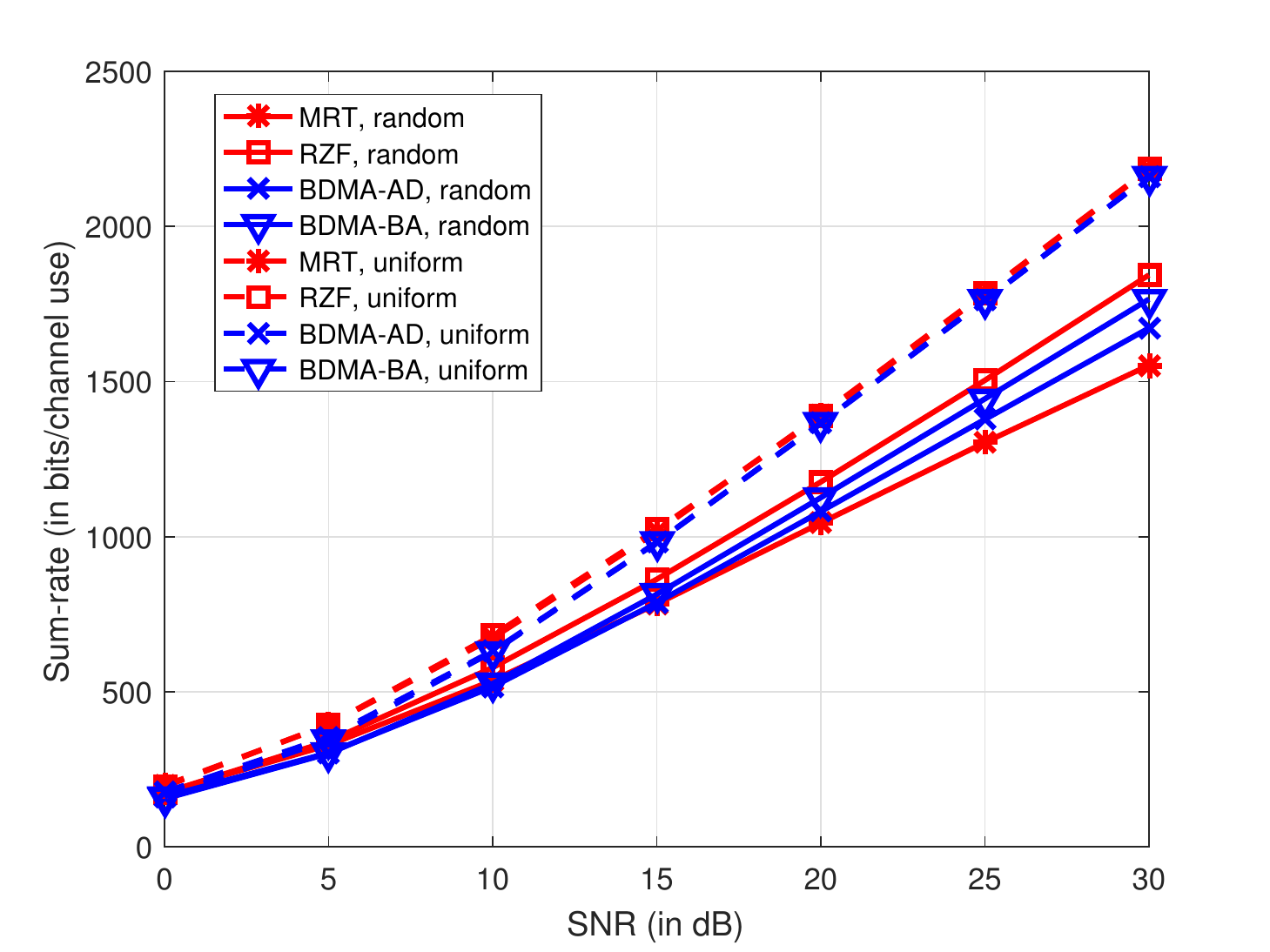}\label{fig:total2}}
\subfigure[]{\includegraphics[width=0.45\textwidth]{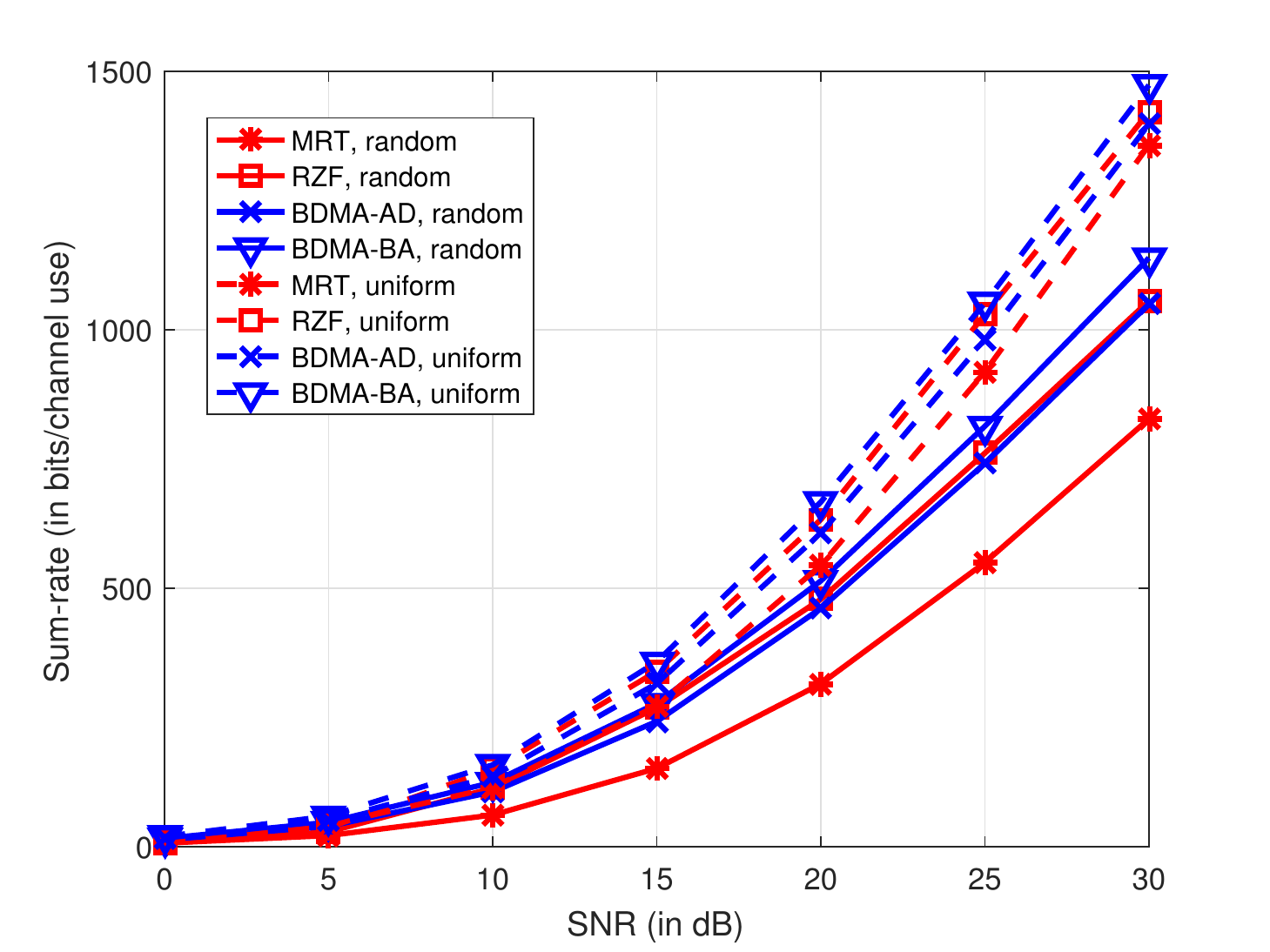}\label{fig:per2}}
\caption{Comparison of sum rate in the wide area scenario, (a) under the total power constraint, (b) under the per LED power constraint.}
\label{fig:5}
\end{figure}

{Fig.~\ref{fig:5} compares the sum rates of different schemes in the wide area scenario. For the uniformly placed case, as UTs are sufficiently separated, there is little interference between UTs. All the transmit schemes have similar performance.
The average rate per UT approaches $4.5$ bits per channel use under the total power constraint, and $3$ bits per channel use under the per LED power constraint.
For the randomly distributed case, 
the interference between UTs increases to degrade the performance,
while the performance gaps between different schemes become larger than those in the uniformly placed case.
Under both power constraints, BDMA-BA has the similar performance with RZF.
}

\begin{figure}[htbp]
\centering
\subfigure[]{\includegraphics[width=0.45\textwidth]{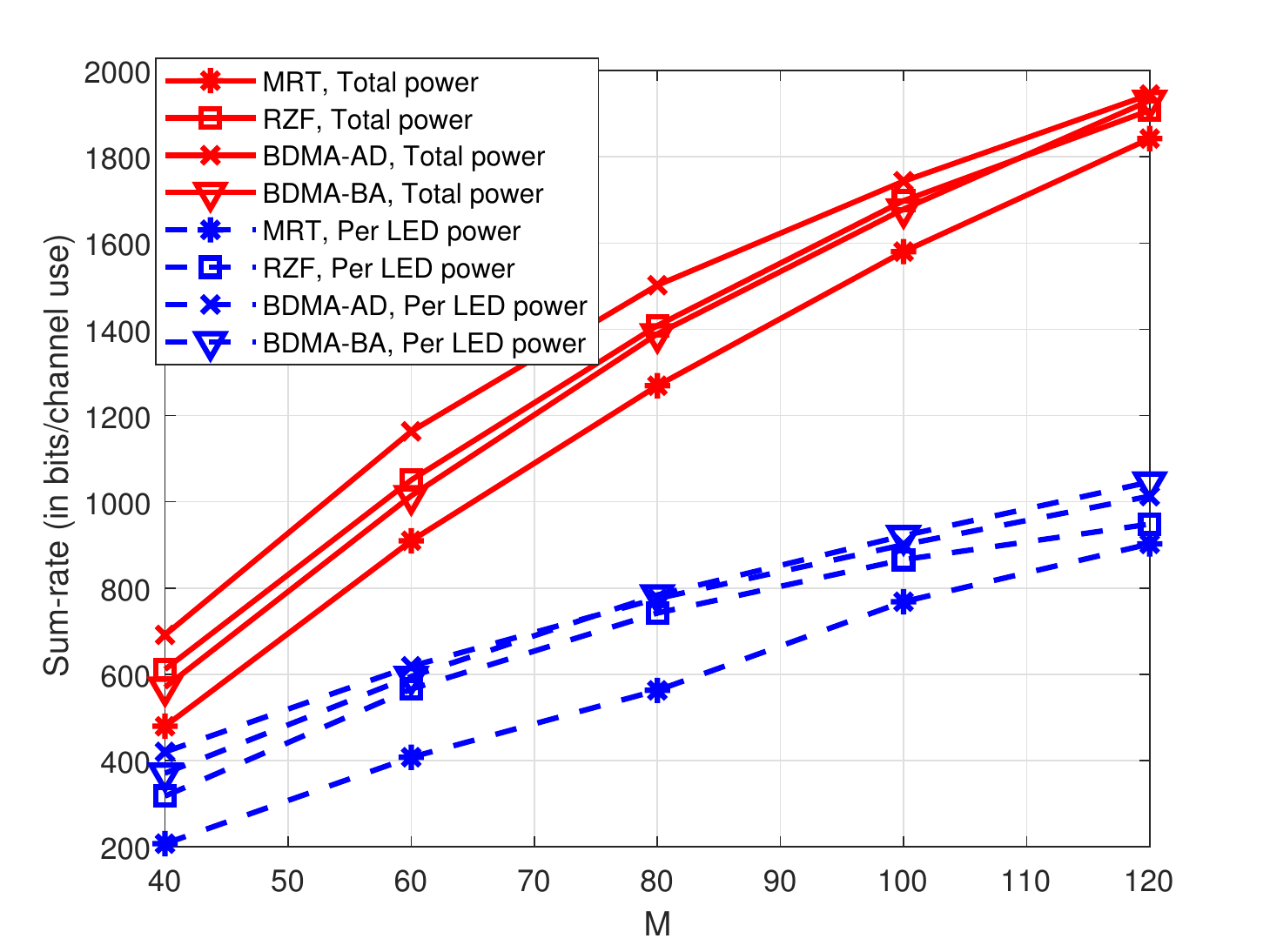}\label{fig:SumRate_N}}
\subfigure[]{\includegraphics[width=0.45\textwidth]{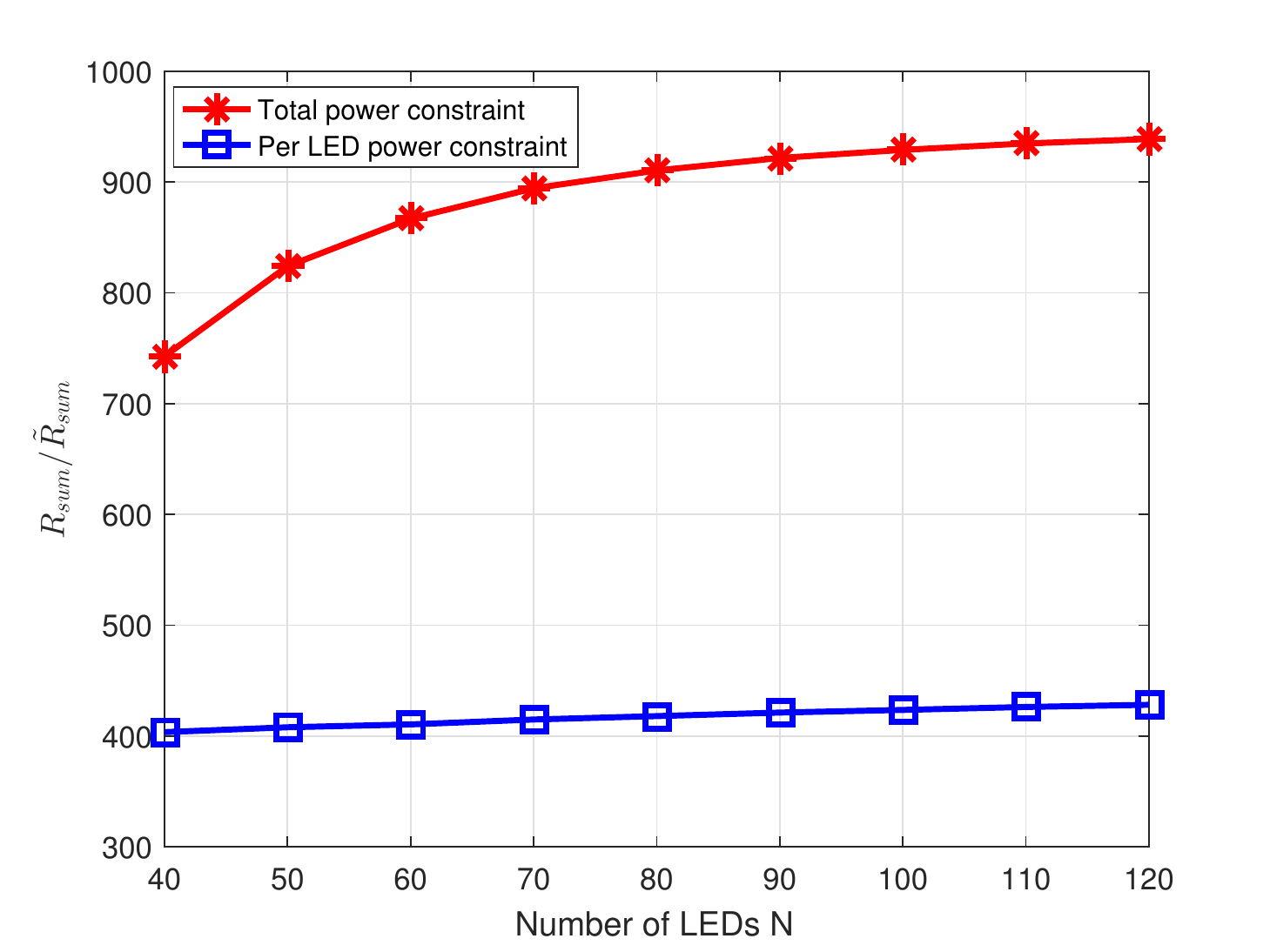}\label{fig:Ratio_N}}
\caption{Comparison of (a) sum rate and (b) sum rate ratio for different $M$.}
\end{figure}

Fig. \ref{fig:SumRate_N} shows the sum rate as the number of transmit LEDs increases. 
Here, we consider {the wide area scenario and} that $K=484$ UTs are randomly distributed.
Under both power constraints, the BDMA-BA can approach the asymptotic sum rate as in \eqref{eq:asymptotic_sum_rate} and \eqref{eq:asymptotic_sum_rate_2}. MRT scheme has a significant performance loss under per LED power constraint.
Recall the theoretic analysis of the sum rate ratio in Section \ref{sub:comparison_of_led_array_without_transmit_lens} and Section \ref{sub:_comparison_with_led_array_without_transmit_lens}.
Fig. \ref{fig:Ratio_N} shows the sum rate ratios as the number of transmit LEDs increases. Compared with the theoretic results, in \eqref{eq:Ratio1}, the ratio for the total power constraint is $2K=968$, and in the figure, 
when $M=70$, the ratio approaches to 900.
Under the per LED power constraint, in \eqref{eq:Ratio2}, the asymptotic ratio is $K=484$, while in the simulation, the ratio is larger than 400.



\section{Conclusion} 
\label{sec:conclusion}

We have investigated beam domain optical wireless massive MIMO communications with a large number of transmitters and a transmit lens equipped at BS. 
{We focused on LED transmitter in this work, and provided a transmit lens based optical channel model. 
With a transmit lens at BS, lights from one LED are refracted to one direction and generate narrow light beams.
With large number of LEDs, the channel vectors for different UTs become asymptotically orthogonal,
and thus, BS has the potential to serve a number of UTs simultaneously. }
For this channel model, we analyzed the performance of MRT/RZF linear precoding, and provided a transmit covariance matrix design.
As the number of LEDs tends to infinity, we designed the transmit covariance matrix  under both total and per LED power constraints. For both power constraints, the optimal transmit policy is to transmit signals to different UTs by non-overlapping beams. Thus, BDMA transmission can achieve the optimal performance. 
Compared with the conventional transmission without a transmit lens, the sum rate of BDMA transmission is improved by $2K$ times under the total power constraint, and $K$ times under the per LED power constraint. 
In addition, for non-asymptotic case, we provided beam allocation algorithms to allocate non-overlapping beams to different UTs.
Simulations illustrate the extremely high spectrum efficiency with massive LEDs and hundreds of UTs.

\appendices

\section{Proof of Theorem \ref{thm:CCCP1}} 
\label{sec:proof_of_theorem_thm:cccp1}

As $\log\det(\cdot)$ function is concave, for the $i$th and $(i+1)$th iteration results, we have
\begin{align}
 	f(\Qc^{(i+1)}) -  g(\Qc^{(i+1)})  
	\ge&  f(\Qc^{(i+1)}) -  g(\Qc^{(i)})  -  \sum_k \tr \left( \left( \frac{\partial}{\partial \Q_k} g(\Qc^{(i)})\right)^T   \left( \Q_k^{(i+1)} - \Q_k^{(i)} \right)  \right)\nn \\
	\ge & f(\Qc^{(i)}) -  g(\Qc^{(i)})  -  \sum_k \tr \left( \left( \frac{\partial}{\partial \Q_k} g(\Qc^{(i)})\right)^T   \left( \Q_k^{(i)} - \Q_k^{(i)} \right)  \right) \nn\\
	= & f(\Qc^{(i)}) -  g(\Qc^{(i)}).
\end{align}
The objective function is monotonic and bounded. Moreover, the set $\Qc$ is closed and bounded. Invoking Theorem 4 in \cite{lanckriet2009convergence}, we have
\begin{align}
    \lim_{i\to\infty} \left ( f(\Qc^{(i)}) - g(\Qc^{(i)}) \right)
    = f(\Qc^{(*)}) - g(\Qc^{(*)}) ,
\end{align}
where $\Qc^{(*)}$ is a generalized fixed point. Then, there exists Lagrange multipliers $\eta^{(*)}$ and $\{\A_{k}^{(*)}\}_{k=1}^{K}$
such that the following KKT conditions hold
\begin{align}\label{eq:KKT}
	&\frac{\partial}{\partial \Q_k} f(\Qc^{(*)}) - \frac{\partial}{\partial \Q_k} g(\Qc^{(*)})
	-\eta^{(*)}\I + \A_k^{(*)} = \bb,  \nn \\
	& \sum_k \tr \left( \Q_k^{(*)} \right)\le P, \eta^{(*)}\ge 0, \eta^{(*)} \left( \sum_k \tr \left( \Q_k^{(*)} \right)- P \right) = 0 , \nn \\
	& \Q_k^{(*)}\succeq \bb, \A_k^{(*)}\succeq \bb, \tr \left( \A_k^{(*)}\Q_k^{(*)} \right)= 0,
\end{align}
which is exactly the KKT condition of problem \eqref{eq:problem1},
and therefore, $\Qc^{(*)}$ is a stationary point of \eqref{eq:problem1}.

{
Moreover, from the KKT conditions, we have
\begin{align}
    \A_k^{(*)} = &\eta^{(*)} \I +  \sum_{i\neq k}\h_i \h_i^T \left(
    (\Rc_i^{(*)})^{-1} -\left( \Rc_i^{(*)} + \gamma\h_i^T  \Q_{i}^{(*)}  \h_i \right)^{-1} \right) \nn\\
   & - \gamma\h_k \h_k^T\left( \Rc_k^{(*)} + \gamma\h_k^T  \Q_{k}^{(*)}  \h_k \right)^{-1},
\end{align}
where $\Rc_k^{(*)} = 1 + \h_k^T \left( \sum_{k'\neq k} \Q_{k'}^{(*)} \right) \h_k$.
To guarantee the semi-definite matrix $\A_k^{(*)}$,
we need $\eta^{(*)} >0$.
The matrix 
\begin{align}
    \eta^{(*)} \I +  \sum_{i\neq k}\h_i \h_i^T \left(
    (\Rc_i^{(*)})^{-1} -\left( \Rc_i^{(*)} + \gamma\h_i^T  \Q_{i}^{(*)}  \h_i \right)^{-1} \right) 
\end{align}
is a positive-definite matrix with rank $M$.
Thus, we have $\rank(\A_k^{(*)} ) \ge M-1$.
Recalling $\A_k^{(*)}\Q_k^{(*)} = \bb$,
the rank of $\Q_k^{(*)}$ satisfies $\rank(\Q_k^{(*)})\le 1$.
}
This completes the proof.
\qed

\section{Proof of Theorem \ref{thm:gamma}} 
\label{sec:proof_of_proposition_pro:1}

The $(m_1,m_2)$th element of $\R_k$ can be expressed as
\begin{align}
    [\R_k]_{m_1m_2} 
    = \left( \frac{A_k M^2 4\Phi_C^2}{d_k^2 \omega^2} \cos(\phi_k) \right)^2 
    I_{i_1j_1} (\psi_{i_1j_1,k})I_{i_2j_2} (\psi_{i_2j_2,k}) ,
\end{align}
where $m_1=(i_1 -1 )M+ j_1$, $m_2 = (i_2-1)M+j_2$.
Then, $\tr(\R_k\Q_k)$ is calculated as
\begin{align}
    \tr(\R_k\Q_k) &= \sum_{m_1}\sum_{m_2} [\R_k]_{m_1m_2} [\Q_k]_{m_2m_1} \nn\\
    &= \sum_{m_1}\sum_{m_2}  \left( \frac{A_k M^2 4\Phi_C^2}{d_k^2 \omega^2} \cos(\phi_k) \right)^2  I_{i_1j_1} (\phi_{i_1j_1,k})I_{i_2j_2} (\phi_{i_2j_2,k}) [\Q_k]_{m_2m_1}. 
\end{align}
Let
\begin{align}
    g_{k} = T_{lens} \frac{A_k 4\Phi_C^2  }{ \omega^2 d_k^2} \frac{(m+1) }{2\pi}  \cos(\phi_k),
\end{align}
and
\begin{align}
    \Ra_{\tsum} =  \frac{1}{2}\sum_k \log \left( 1 + \gamma \frac{M^4g_k^2 [\Q_k]_{m_km_k}}{1+M^4 g_k^2\sum_{k'\neq k} [\Q_{k'}]_{m_km_k}} \right).
\end{align}
As the number of LED increases, there exists LED $(i_k, j_k)$ satisfying $\phi_{k,i_k j_k} = 0$.
Thus, we have
\begin{align}\label{eq:B1}
    \lim_{M\to \infty} \left( R_{\tsum}-\Ra_{\tsum} \right) =&\frac{1}{2}\sum_k  \lim_{M\to \infty}
    \log \left( \frac{1+  \tr \left( \R_k \sum_{k'}\Q_{k'} \right)-(1- \gamma)\tr \left( \R_k \Q_{k} \right)  }{1+  M^4 g_k^2\sum_{k'} [\Q_{k'}]_{m_km_k}-(1- \gamma)M^4 g_k^2[\Q_{k}]_{m_km_k}   }\right) \nn\\
    & + \frac{1}{2}\sum_k  \lim_{M\to \infty}
    \log \left(  \frac{1+M^4 g_k^2\sum_{k'\neq k} [\Q_{k'}]_{m_km_k}}{1 + \tr \left( \R_k \sum_{k'\neq k}\Q_{k'} \right)} \right).
\end{align}
Consider the limit of the first term in \eqref{eq:B1}.
For the case of $\sum_{k'} [\Q_{k'}]_{m_km_k}=0$, we have
\begin{align}\label{eq:B2}
    &\lim_{M\to \infty} \frac{1+  \tr \left( \R_k \sum_{k'}\Q_{k'} \right)-(1- \gamma)\tr \left( \R_k \Q_{k} \right)  }{1+  M^4 g_k^2\sum_{k'} [\Q_{k'}]_{m_km_k}-(1- \gamma)M^4 g_k^2[\Q_{k}]_{m_km_k}   } \nn\\
    & = 1+ \lim_{M\to \infty} \tr \left( \R_k \sum_{k'}\Q_{k'} \right)-(1- \gamma)\tr \left( \R_k \Q_{k} \right) \nn\\
    & = 1.
\end{align}

For the case of $\sum_{k'} [\Q_{k'}]_{m_km_k}\neq 0$, we have
\begin{align}\label{eq:B3}
    &\lim_{M\to \infty} \frac{1+  \tr \left( \R_k \sum_{k'}\Q_{k'} \right)-(1- \gamma)\tr \left( \R_k \Q_{k} \right)  }{1+  M^4 g_k^2\sum_{k'} [\Q_{k'}]_{m_km_k}-(1- \gamma)M^4 g_k^2[\Q_{k}]_{m_km_k}   } \nn\\
    &=\frac{\frac{1}{M^4}+  \frac{1}{M^4}\tr \left( \R_k \sum_{k'}\Q_{k'} \right)-(1- \gamma)\frac{1}{M^4}\tr \left( \R_k \Q_{k} \right)  }{\frac{1}{M^4}+  g_k^2\sum_{k'} [\Q_{k'}]_{m_km_k}-(1- \gamma) g_k^2[\Q_{k}]_{m_km_k}   }. 
\end{align}
The limit of $\frac{1}{M^4} \tr \left( \R_k \sum_{k'}\Q_{k'} \right)$ exists, and can be calculated as
\begin{align}
    &\lim_{M\to \infty}\frac{ 1}{M^4} \tr \left( \R_k \sum_{k'}\Q_{k'} \right)\nn\\
    &=\sum_{k'}\lim_{M\to \infty} \frac{1}{M^4}\sum_{m_1}\sum_{m_2}  \left( T_{lens}\frac{A_k M^2 4\Phi_C^2}{d_k^2 \omega^2} \cos(\phi_k) \right)^2  I_{i_1j_1} (\phi_{i_1j_1,k})I_{i_2j_2} (\phi_{i_2j_2,k}) [\Q_{k'}]_{m_2m_1} \nn\\
    &=\sum_{k'}\lim_{M\to \infty}   \left(T_{lens} \frac{A_k  4\Phi_C^2}{d_k^2 \omega^2}\frac{m+1}{2\pi} \cos(\phi_k) \right)^2  [\Q_{k'}]_{m_km_k}\nn\\
    &= \sum_{k'}  g_k^2[\Q_{k'}]_{m_km_k} .
\end{align}
Then, the limit \eqref{eq:B3} is 
\begin{align}
    \lim_{M\to \infty} \frac{1+  \tr \left( \R_k \sum_{k'}\Q_{k'} \right)-(1- \gamma)\tr \left( \R_k \Q_{k} \right)  }{1+  M^4 g_k^2\sum_{k'} [\Q_{k'}]_{m_km_k}-(1- \gamma)M^4 g_k^2[\Q_{k}]_{m_km_k}   } =1.
\end{align}
Similarly, the limit of the second term in $\log$ operation in \eqref{eq:B1} can be calculated as
\begin{align}
    \lim_{M\to \infty} \frac{1+M^4 g_k^2\sum_{k'\neq k} [\Q_{k'}]_{m_km_k}}{1 + \tr \left( \R_k \sum_{k'\neq k}\Q_{k'} \right)}=1 .
\end{align}
As $\log$ function is continuous and $\log(1)=0$, the limit of $R_{\tsum}-\Ra_{\tsum}$ is 
\begin{align}
    \lim_{N\to \infty} \left( R_{\tsum}-\Ra_{\tsum} \right) = 0.
\end{align}
This completes the proof.
\qed

\section{Proof of Theorem \ref{thm:4}}\label{proof:thm4}

From \eqref{eq:asymp_sum_rate},
the asymptotic sum rate $\Ra_{\tsum}$ depends on the diagonal elements of $\Q_k$.
Moreover, as UTs are in different positions, 
different UTs receive signals from different LEDs, i.e., $m_k\neq m_{k'}$, for $k\neq k'$.
The optimal transmit covariance matrix $\Q_{k}$ should satisfy
\begin{align}
    \left[ \Q_k \right]_{m'_k m'_k} = 0, \quad m'_k \neq m_k.
\end{align}
Since only one diagonal element of $\Q_k$ is non-zero,
and $\Q_k$ is a positive-semidefinite matrix,
$\Q_k$ must be a diagonal matrix.
Under this condition, the problem \eqref{eq:Prob5_1} is reduced to 
\begin{align}
    \max_{\Q_1,\Q_2,\cdots,\Q_K}  &  \frac{1}{2}\sum_k \log \left( 1 +  \gamma M^4 g_k^2  [\Q_k]_{m_km_k}  \right) \nn\\
    \tst \quad & \sum_k [\Q_k]_{m_km_k}   \le {P} ,\quad [\Q_k]_{m_km_k} \ge 0.
\end{align}
For this problem, we can have the water-filling result:
\begin{align}\label{eq:water_filling}
    [\Q_k]_{m_km_k} = \left(  \frac{1}{\nu} - \frac{1}{ \gamma M^4 g_k^2} \right)^+ ,
\end{align}
where $(x)^+=\max \{x,0\}$, $\nu$ is the Lagrange multiplier with the condition
\begin{align}
    \sum_m \left(  \frac{1}{\nu} - \frac{1}{ \gamma M^4 g_k^2} \right)^+ = \bar{P} . 
\end{align}
According to Theorem \ref{thm:gamma}, the asymptotically optimal sum rate can be expressed as
\begin{align}
    R_{\tsum}^{o} - \frac{1}{2} \sum_k\log \left( 1 +  \gamma M^4 g_k^2  [\Q_k]_{m_km_k}  \right)\to 0.
\end{align}
This completes the proof.
\qed

\section{Proof of Theorem \ref{thm:5}}\label{proof:thm5}

Let $\Lambdam_k = \Lambdam  \B_k$, where $\B_k$ is an auxiliary diagonal matrix satisfying $\sum_k \B_k = \I$ and $\B_k \succeq \bb$.
The optimization problem \eqref{eq:natp1} is equal to
\begin{align}
    \max_{\Lambdam} \max_{\B_1,\B_2,\ldots,\B_K}
    &\frac{1}{2}\sum_k ( \log(1\!+\tr(\R_k \Lambdam)\!-\!(1\!- \!\gamma)\tr(\R_k \Lambdam  \B_k))
    \!-\! \log(1+\tr(\R_k \Lambdam)\!-\!\tr(\R_k \Lambdam  \B_k)) ) \nn\\
    \tst \quad  & \tr (\Lambdam)\le P, \quad \Lambdam \succeq \bb, \\
    & \sum_k \B_k = \I , \quad \B_k \succeq \bb.
\end{align}
For any fixed $\Lambdam$, 
let $a_k = 1+\tr(\R_k \Lambdam)$, and
consider the inner optimization problem on $\B_k$, which is given by
\begin{align}\label{eq:96}
    \max_{\B_1,\B_2,\ldots,\B_K}
    &\sum_k ( \log(a_k-(1- \gamma)\tr(\R_k \Lambdam \B_k))
    - \log(a_k-\tr(\R_k \Lambdam \B_k)) ) \nn\\
    \tst \quad  &  \sum_k \B_k = \I , \quad \B_k \succeq \bb.
\end{align}
Define $\Rt_k = \R_k \odot \I$, where $\Rt_k$ is a diagonal matrix, which consists of the diagonal elements of $\R_k$.
As $\Lambdam$ and $\B_k$ are diagonal matrices, we have
$\tr( \R_k \Lambdam \B_{k} ) = \tr( \Rt_k \Lambdam \B_{k} )$.

Let diagonal matrices $\A_k$ and $\C$ be Lagrange multipliers, and we have the cost function as
\begin{align}
	\Lc = &\sum_k ( \log(a_k-(1- \gamma)\tr(\Rt_k \Lambdam \B_k))
    - \log(a_k-\tr(\Rt_k \Lambdam \B_k)) ) \nn\\
    &+ \tr \left(  \C \left( \sum_k \B_k - \I \right) \right)- \tr \left( \A_k\B_k \right).
\end{align}
The Karush-Kuhn-Tucker (KKT) conditions for optimum $\B_k$, $\A_k$ and $\C$ can be written as
\begin{align}
	&\frac{\partial}{\partial \B_k}\Lc = (1- \gamma) \!\left(a_k\! - \! (1- \gamma)\tr(\Rt_k \Lambdam \B_k)\right) ^{-1}\!\Rt_k \Lambdam \! +  \!
    \left( a_k\! -\tr(\Rt_k \Lambdam \B_k)\right) ^{-1}\! \Rt_k \Lambdam \! +\! \C\! -\! \A_k = \bb,  \nn\\
	& \sum_k \B_k - \I  = \bb , \quad \A_k\B_k = \bb, \quad \B_k \succeq \bb, \quad \A_k \succeq \bb .
\end{align}
Consider the KKT conditions for UT ${k_1}$ and UT $k_2$, and we have
\begin{align}
	\C & = \A_{k_1}\! -(1- \gamma) \!\left(a_{k_1}\! - \! (1- \gamma)\tr(\Rt_{k_1} \Lambdam \B_{k_1})\right) ^{-1}\!\Rt_{k_1} \Lambdam \! -  \!
    \left( a_{k_1}\! -\tr(\Rt_{k_1} \Lambdam \B_{k_1})\right) ^{-1}\! \Rt_{k_1} \Lambdam , \nn\\
	\C & = \A_{k_2}\! - (1- \gamma) \!\left(a_{k_2}\! - \! (1- \gamma)\tr(\Rt_{k_2} \Lambdam \B_{k_2})\right) ^{-1}\!\Rt_{k_2} \Lambdam \! +  \!
    \left( a_{k_2}\! -\tr(\Rt_{k_2} \Lambdam \B_{k_2})\right) ^{-1}\! \Rt_{k_2} \Lambdam .
\end{align}
For the $m$th diagonal elements, if $[\B_{{k_1}}]_{mm}\neq 0$ and $[\B_{{k_2}}]_{mm}\neq 0$, there exists $[\A_{{k_1}}]_{mm}= 0$ and $[\A_{{k_2}}]_{mm}= 0$.
Thus, we have
\begin{align}\label{eq:max_point}
	[\C]_{mm} &=\!(1- \!\gamma) \!\left(a_{k_1}\! - \! (1-\! \gamma)\tr(\Rt_{k_1} \Lambdam \B_{k_1})\right) ^{-1}\!\!\left[ \Rt_{k_1} \Lambdam \right]_{mm} \! \!\!\!-  \!
    \left( a_{k_1}\! -\tr(\Rt_{k_1} \Lambdam \B_{k_1})\right) ^{-1}\!\! \left[ \Rt_{k_1} \Lambdam \right]_{mm} \nn\\
	&=\!(1- \!\gamma) \!\left(a_{k_2}\! - \! (1-\! \gamma)\tr(\Rt_{k_2} \Lambdam \B_{k_2})\right) ^{-1}\!\!\left[ \Rt_{k_2} \Lambdam \right]_{mm} \! \!\!\!-  \!
    \left( a_{k_2}\! -\tr(\Rt_{k_2} \Lambdam \B_{k_2})\right) ^{-1}\!\! \left[ \Rt_{k_2} \Lambdam \right]_{mm}.
\end{align}
From Lemma 2 in \cite{BDMA}, the objective function in \eqref{eq:96} is matrix convex on $\B_k$.\
Thus, the solution of equation \eqref{eq:max_point} is a minimum point.
Therefore, the solution of problem \eqref{eq:96} should satisfies
$[\B_{{k_1}}]_{mm} [\B_{{k_2}}]_{mm} = 0$.
This means that for different UTs $k_1$ and $k_2$, we have
$\B_{k_1} \B_{k_2} = \bb$.
Therefore we have \eqref{eq:orthogonal_conditions}. This completes the proof.
\qed

\bibliographystyle{IEEEtran}
\bibliography{IEEEabrv,mybibfile}
\end{document}